\documentclass[a4paper,11pt]{article}
\usepackage{jcappub} 
\usepackage{lineno}

\newcommand{\I}{(\mathcal{I})}

\title{\boldmath Exploring the interplay of late-time dynamical dark energy and new physics before recombination}

\author{Alex Gonz\'alez-Fuentes}
\author{and Adri\`a G\'omez-Valent}
\affiliation{Departament de F\'isica Qu\`antica i Astrof\'isica (FQA) and Institut
de Ciències del Cosmos (ICCUB), Universitat de Barcelona (UB), c. Martí i Franqués 1, 08028 Barcelona, Catalonia, Spain}

\emailAdd{agonzalezfuentes@icc.ub.edu}
\emailAdd{agomezvalent@icc.ub.edu}

\abstract{Cosmological models in which the effective dark energy (DE) equation-of-state parameter crosses the phantom divide at 
$z_c \sim 0.3-0.6$, while exhibiting quintessence-like behavior at low redshift, provide a significantly improved fit to current datasets, yielding indications of late-time DE dynamics at the $\sim 3\sigma$ CL. However, such scenarios continue to favor low values of the Hubble constant, remaining in tension with the SH0ES distance-ladder determination. In the absence of unaccounted-for systematic effects, this tension may point to the presence of new physics operating prior to the decoupling of CMB photons. In this work, we reconstruct the relevant background DE functions using the Weighted Function Regression (WFR) method, introducing three main improvements with respect to our previous analysis in JCAP 12 (2025) 049. First, we adopt the Frequentist--Bayesian approach for determining the WFR weights, and demonstrate full consistency with results obtained using information criteria. Second, we combine state-of-the-art CMB and BAO measurements with the newly recalibrated DES supernova sample (DES-Dovekie), and compare our findings with those derived from Pantheon+, still assuming standard pre-recombination physics. Third, we study in a model-independent manner the viability of the early-time ``solutions'' to the Hubble tension and how they affect the evidence for dynamical DE at late times, under the influence of the SH0ES and the more conservative CCHP calibration of the cosmic ladders, separately. We find that, if the physics prior to decoupling is unmodified, the probability of phantom crossing is $\sim 96.7\text{--}98.5\%$, with $\Lambda$CDM excluded at $\sim 2.5\sigma$ and $\sim 3\sigma$ CL for the models with the largest weights when using Pantheon+ and DES-Dovekie, respectively. New physics before recombination can alleviate the Hubble tension, but requires extremely large values of the reduced matter density parameter $\omega_m$ when the SH0ES calibration is employed, in strong tension with those inferred from full CMB analyses. This raises serious concerns about the actual viability of these models to explain the SH0ES measurement. We find that phantom crossing, while not excluded, is no longer required, with only a very mild preference for quintessence. Nevertheless, given the aforementioned tension in $\omega_m$, it would be rash to draw firm conclusions about how the dynamical DE signal is affected in these scenarios.}


\begin{document}
\maketitle
\flushbottom

\section{Introduction}
\label{sec:intro}

Recent cosmological data from several probes -- namely the cosmic microwave background (CMB) \cite{Planck:2018vyg,Efstathiou:2019mdh,Rosenberg:2022sdy}, Type Ia supernovae (SNIa) \cite{Scolnic:2021amr,Rubin:2023jdq,DES:2024hip,DES:2024jxu,DES:2025sig}, and baryon acoustic oscillations (BAO) \cite{eBOSS:2020yzd,DESI:2025zgx} -- may be hinting at a more complex dark sector than that encountered in the standard $\Lambda$CDM model. In particular, recent analyses suggest that the effective\footnote{Throughout this work, by {\it effective} dark energy we refer to a component that encapsulates any late-universe physics beyond $\Lambda$CDM that is not accounted for by the degrees of freedom of the Standard Model of particle physics, non-interacting cold dark matter, or a cosmological constant. This can refer, e.g., to slow-rolling scalar fields minimally coupled to gravity \cite{Copeland:2006wr} or running vacuum scenarios \cite{Mavromatos:2020kzj,SolaPeracaula:2022hpd,Moreno-Pulido:2022phq,Moreno-Pulido:2022upl,SolaPeracaula:2026trz}. The effective DE sector may also encompass possible modified-gravity effects \cite{Clifton:2011jh}. See also \cite{Amendola:2015ksp}.} dark energy (DE) component is not only dynamical, but may exhibit a peak in its energy density within the redshift range $z_c \sim 0.3\text{--}0.6$. Such a feature would imply an effective crossing of the phantom divide, with DE behaving as phantom ($w<-1$) at higher redshifts and as quintessence-like ($w>-1$) at lower redshifts, giving rise to the so-called quintom scenario (see  the reviews \cite{Cai:2009zp,Cai:2025mas}\footnote{To the best of our knowledge, the first quintom scenario was proposed in \cite{Feng:2004ad}, with the crossing happening in the opposite direction, from quintessence in the past to phantom at present. See also \cite{Grande:2006nn,Grande:2006qi} for an early realization of the correct quintom phenomenology (from phantom to quintessence) in the context of the $\Lambda$XCDM model.}). This would require a more complex dynamical behavior than that suggested by previous indications in the literature \cite{Alam:2003fg,Alam:2004jy,Salvatelli:2014zta,Sahni:2014ooa,Sola:2015wwa,Sola:2016jky,SolaPeracaula:2016qlq,Zhao:2017cud,SolaPeracaula:2017esw,Sola:2017znb,SolaPeracaula:2018wwm}. These hints have been detected both in analyses relying on concrete models and parametrizations  (see, e.g., \cite{Chakraborty:2024xas,Gomez-Valent:2024tdb,Ye:2024ywg,Wolf:2024eph,RoyChoudhury:2024wri,Park:2024pew,Gomez-Valent:2024ejh,Odintsov:2024woi,Giare:2025pzu,Wolf:2025jlc,Silva:2025hxw,Chakraborty:2025syu,Pan:2025psn,Khoury:2025txd,Wolf:2025jed,RoyChoudhury:2025dhe,DESI:2025wyn,Scherer:2025esj,Yang:2025mws,Chen:2025wwn,Giani:2025hhs,Nojiri:2025low,Mishra:2025goj,Braglia:2025gdo,Asorey:2025hgx,Gomez-Valent:2025mfl,RoyChoudhury:2025iis,Sohail:2025mma,Yao:2025wlx,Wang:2025znm,Goh:2025upc,Tsujikawa:2025wca,Wolf:2025acj,SanchezLopez:2025uzw,Efstratiou:2025iqi,Camarena:2025upt,Artola:2025zzb,Li:2025vuh,Ghedini:2025epp,deCruzPerez:2025dni,Cheng:2025yue,Li:2026xaz,Xu:2026sbw,Ibarra-Uriondo:2026zbp,Akarsu:2026anp,Park:2026iqa,Wang:2026wrk,Gomez-Valent:2026ept,Artola:2026tgs,Toda:2026yum}) and also in studies aiming to decrease the impact of possible model-dependent biases through the agnostic reconstruction of the most relevant cosmological observables \cite{DESI:2024aqx,DESI:2025fii,Gonzalez-Fuentes:2025lei} (see also \cite{Jiang:2024xnu,Berti:2025phi}).

However, even within a given cosmological model or DE parametrization, and for a fixed choice of dataset, the inferred statistical significance of a phantom-divide crossing and of DE dynamics depends on the statistical method used to quantify departures from $\Lambda$CDM. For example, there is no full consensus about the actual significance of the evidence for DE dynamics obtained with the Chevallier-Polarski-Linder (CPL) -- also known as $w_0w_a$CDM -- parametrization of the DE equation-of-state (EoS) parameter, $w_{\rm DE}(a)=w_0+w_a(1-a)$ \cite{Chevallier:2000qy,Linder:2002et}. While likelihood-ratio tests based on CMB+SNIa+BAO fits including DESI data lead to an exclusion of $\Lambda$CDM at the $\sim 3$--$4\sigma$ confidence level (CL) \cite{DESI:2025zgx,DES:2025sig}\footnote{See also \cite{Park:2024vrw,Gomez-Valent:2024ejh,Lu:2025gki} for similar results obtained with BAO measurements from eBOSS and other pre-DESI galaxy surveys.}, in agreement with the contour plots in the $w_0$--$w_a$ plane and profile likelihood analyses \cite{Herold:2025hkb}, Bayesian model-comparison studies based on exact Bayes factors and Jeffreys' scale \cite{Trotta:2008qt} appear to suggest a markedly different conclusion. In particular, these analyses find no statistically significant evidence in favor of departures from the standard $\Lambda$CDM scenario \cite{Ong:2025utx,Hergt:2026moc,Ong:2026tta}. Nevertheless, it is important to recall that the computation of Bayesian evidences and Bayes factors depends sensitively on the choice of priors \cite{Efstathiou:2008ed}. This becomes particularly problematic when considering phenomenological parametrizations such as the CPL model, for which truly robust and physically well-motivated priors are difficult to define in a precise way. In this context, the prior choice is inevitably to some extent subjective.

\begin{table}
    \centering
    \begin{tabular}{cc}
    \hline
       $\beta\equiv 2\ln B_{IJ}$ & Evidence for model $I$ versus model $J$ \\ \hline
        $\leq 2$ & Inconclusive \\ 
        $2\leq \beta < 5$ & Weak \\ 
        $5 \leq \beta < 10$ & Moderate \\ 
        $\geq 10$ & Strong \\ \hline
    \end{tabular}
    \caption{Revised Jeffreys' scale from \cite{Trotta:2008qt}, with the Bayes factor $B_{IJ}=\mathcal{E}_I/\mathcal{E}_J$ and $\mathcal{E}_I$ the Bayesian evidence of model $I$.}
    \label{tab:jeffreysscale}
\end{table}

Even when adopting nominally uninformative priors -- with a prior volume $\mathcal{V}$ much larger than the region of parameter space favored by the likelihood, so that the posterior shape remains essentially unaffected -- the use of an even broader prior volume, e.g.\ $2\mathcal{V}$, can still have a non-negligible impact on the Bayesian evidence and hence on the model-selection outcome. This strong prior-volume dependence can appear counterintuitive and deserves, at the very least, a detailed investigation of how Bayes factors vary with the assumed prior volume (see, e.g., \cite{SolaPeracaula:2018wwm,Patel:2024odo}). Nevertheless, in order to extract quantitative conclusions from a Bayesian model-comparison analysis, one must ultimately adopt a specific prior choice and interpret the resulting Bayes factor through a reference scale that assigns a level of statistical support. A commonly used criterion is the Jeffreys' scale \cite{Jeffreys1961}, or its revised version \cite{Trotta:2008qt, Trotta:2005ar} (cf. Table \ref{tab:jeffreysscale}). However, it should be emphasized that this scale cannot be regarded as universal, and its applicability may depend on the context and the class of models and datasets under consideration \cite{Nesseris:2012cq}.

A possible resolution to these two issues -- namely, the undesired impact of uninformative priors on Bayesian model-comparison analyses and the potential biases associated with the use of Jeffreys' scale --  has been proposed in Refs. \cite{Jenkins_2011,Keeley:2021dmx,Amendola:2024prl}. In this approach, the Bayes factor is treated as a frequentist random variable, and a $p-$value is assigned to the ``observed'' Bayes factor. This allows one to quantify the statistical significance of the result without relying on Jeffreys' scale and with very mild or null sensitivity on the priors. In Ref.~\cite{Amendola:2024prl}, this framework is referred to as the Frequentist--Bayesian (FB) method (see \cite{Sakr:2025chr} for a recent application of it).
 
In this work, we first use the FB method to determine the evidence for dynamical DE in the context of the CPL parametrization, using state-of-the-art cosmological data, including the re-calibrated SNIa from the Dark Energy Survey (DES) contained in the so-called DES-Dovekie sample \cite{DES:2025sig}. Moreover, we show that the inferred statistical evidence with respect to the $\Lambda$CDM derived from the FB method coincides with the one obtained with the likelihood-ratio test when the posterior distribution is Gaussian, as it is the case with the CPL parametrization.  

Following this preliminary step, we employ the Weighted Function Regression (WFR) method to reconstruct the effective dark energy EoS parameter, the corresponding DE density, and other relevant background functions. The main goal is to reduce the impact of potential parametrization-dependent features that could bias our conclusions. The WFR approach has been applied in a variety of cosmological contexts \cite{Liddle:2006kn,Gomez-Valent:2018hwc,Gomez-Valent:2018gvm,Gonzalez-Fuentes:2025lei}. The method relies on a set of basis functions -- with the basis being formally infinite --, each characterized by its own fitting performance, and combines them into a final reconstruction by assigning Bayesian weights to their individual contributions, which reward their ability to describe the data and duly penalize the use of additional degrees of freedom. This analysis can be regarded as a follow-up to our previous work \cite{Gonzalez-Fuentes:2025lei}. In addition to incorporating the re-calibrated DES SNIa sample -- which corrects issues affecting the earlier DES-Y5 dataset \cite{DES:2024hip,DES:2024jxu} -- we also improve a key ingredient of the reconstruction pipeline: the computation of the Bayesian weights. Rather than relying on the Akaike and Deviance Information Criteria (AIC \cite{Akaike} and DIC \cite{DIC}), we adopt the FB method and show that the resulting WFR weights are fully consistent with those obtained from AIC and DIC. This analysis was deferred to future work in \cite{Gonzalez-Fuentes:2025lei}, and we address it here.

Finally, we extend the analysis of \cite{Gonzalez-Fuentes:2025lei} by studying in detail the interplay between dynamical dark energy in the late universe and possible new physics operating before recombination. In particular, we pay special attention to the impact of such early-time modifications -- which may be required to resolve the Hubble tension -- on the statistical significance of the evidence for dynamical DE at low redshifts. It is very well-known that by only considering dynamical DE at low redshifts we cannot ameliorate the status of the $H_0$ tension, at least if one considers standard anisotropic BAO data in the fitting analyses \cite{Sola:2017znb,Knox:2019rjx,Krishnan:2021dyb,Lee:2022cyh,Gomez-Valent:2023uof,Pedrotti:2025ccw,Bansal:2026axl}. The reason is that addressing the Hubble tension without invoking new physics before recombination would typically lead to a mismatch between the cosmological distances inferred from calibrated BAO and SNIa observations once they are linked via the Etherington distance-duality relation, see, e.g.,  \cite{Poulin:2024ken, Bousis:2024rnb}. Other possibilities that have been entertained in the literature -- which we will not consider here -- are an ultra-late-time transition in the absolute magnitude of SNIa happening between the second rung of the direct cosmic distance ladder and the Hubble flow, probably induced by some gravitational transition   \cite{Marra:2021fvf,Alestas:2021luu,Gomez-Valent:2023uof,Ruchika:2024ymt,Bansal:2026axl}, or violations of the Etherington relation \cite{Teixeira:2025czm}. Indications for the latter are absent in uncalibrated SNIa and BAO data \cite{Favale:2024sdq}, but become important when the ladders are calibrated using the SH0ES measurements assuming standard physics before recombination\footnote{The need of physics beyond $\Lambda$CDM to address the Hubble tension and other anomalies can manifest itself as an effective running of $H_0$, see, e.g.,  \cite{Krishnan:2020vaf,Dainotti:2021pqg,Dainotti:2022bzg}.}. We refer the reader to dedicated reviews on the Hubble tension for further details  \cite{Perivolaropoulos:2021jda,Freedman:2024eph,CosmoVerseNetwork:2025alb,H0DN:2025lyy}.

In this work, we investigate in a model-agnostic way whether the presence of new physics prior to the decoupling of CMB photons modifies the evidence for dynamical dark energy at low redshifts, assuming that the $H_0$ tension is real, i.e., that it is not caused by unaccounted-for systematics in the data. This model independence is reflected both in our treatment of late-time DE dynamics -- for which we employ the WFR reconstruction technique, thus avoiding the assumption of a specific parametrization -- and in our approach to different pre-recombination scenarios. 

There have been several recent works aiming to study the interplay of new physics at late and early times, which are worth commenting on briefly in order to better highlight the novel features of our study. The work \cite{Chaussidon:2025npr} analyzes how the fitting performance of an ultra-light axion-like early dark energy (EDE) component, combined with standard $\Lambda$CDM physics at low redshifts, compares with that of the CPL parametrization. While EDE can lead to a substantial improvement in the goodness of fit to the CMB+BAO+SNIa data with respect to $\Lambda$CDM, it is not able to compete with the CPL, yielding a minimum $\chi^2$ value that is $11.5$ units higher than in the CPL case (cf. their Table IV and Fig. 6). However, EDE is able to produce higher values of $H_0$. This may point to the need for combining dynamical DE with new physics before recombination in order to address both the Hubble tension (see also \cite{Vagnozzi:2023nrq}) and the tension between CMB and low-$z$ data, although in these cases the evidence for DE dynamics is less significant than in the absence of EDE \cite{Wang:2024dka,Pang:2025lvh,Smith:2025icl}.

The author of Ref. \cite{Adi:2025hyj} investigates evidence for dynamical dark energy within the CPL parametrization, and accounts for new physics before recombination in a more model-independent fashion by adopting a prior on $r_d$ designed to yield high values of $H_0$. The analysis considers only BAO+SNIa data, together with a SH0ES prior on $H_0$ and a BBN prior on $\omega_b$. As the author himself acknowledges, this approach does not generally produce cosmological expansion histories compatible with the CMB, in particular with the angular diameter distance to the last-scattering surface inferred from precise measurements of the acoustic scale $\theta_*$. In contrast, we include CMB distance priors to ensure that the expansion histories obtained in our analysis satisfy the most basic CMB constraints (see \cite{Weiner:2026sfm, Shlivko:2026jxa} for recent studies on CMB-BAO tension in multiple scenarios and \cite{Pedrotti:2026dwj} for a model-independent background reconstruction of the pre-recombination expansion, with the necessary increase in $H(z)$).

Given the wide range of theoretical frameworks proposed in the literature, we adopt a minimalist methodology to construct a setup that allows us to extract robust conclusions valid for any cosmology that modifies the physics prior to decoupling. Our treatment of the pre-recombination physics is much closer in philosophy to that of Ref. \cite{Sharma:2025iux} (see also \cite{Lin:2021sfs,GarciaEscudero:2025lef, Wang:2025mqz}). Although all early-universe resolutions of the Hubble tension require a reduction of the comoving sound horizon at both the baryon-drag ($r_d\equiv r_s(z_d)$) and photon-decoupling ($r_*\equiv r_s(z_*)$) epochs, the ratio of these scales remains remarkably stable across models \cite{Lin:2021sfs,Sharma:2025iux,Wang:2025mqz}. We exploit this fact in our analysis, fitting directly $r_d$ and considering a prior on the precise measurement of the CMB acoustic scale $\theta_*$ in combination with low-$z$ BAO and SNIa data, a BBN prior on $\omega_b$ and the distance-ladder calibration from a couple of sources, including SH0ES \cite{Scolnic:2021amr}. Although the value of $z_*$ can vary significantly depending on the underlying theoretical setup\footnote{Such deviations can be non-negligible in some scenarios, for example in models with primordial magnetic fields \cite{Jedamzik:2020krr,Mirpoorian:2024fka,Mirpoorian:2025rfp}. In these cases, CMB photon decoupling occurs slightly earlier than in $\Lambda$CDM and in other early-time solutions such as EDE or early modified gravity models (see e.g. \cite{Poulin:2018cxd,Agrawal:2019lmo,SolaPeracaula:2019zsl,Niedermann:2020dwg,FrancoAbellan:2023gec}).}, the distance to the last-scattering surface is largely insensitive to its value and this is why our conclusions will apply for a very diverse spectrum of models. We exploit this framework to assess how the statistical evidence for late-time dynamical DE is affected by different classes of new physics operating before recombination. This is a very timely and well-motivated question. As we will show, any early-time modification required to resolve the Hubble tension leads to very large values of the reduced matter density parameter $\omega_m$, in clear tension with those inferred from full CMB analyses. While this was already noted in the very exhaustive work \cite{Poulin:2024ken} in the framework  of several pre-recombination modifications, considering the CPL and non-zero curvature at low redshifts (the authors of \cite{Pedrotti:2024kpn} reached similar conclusions, but assuming $\Lambda$CDM in the late universe), we emphasize that this might be an important bottleneck for these models. This issue raises serious concerns about their viability or, at the very least, their ability to efficiently resolve the $H_0$ tension. Explaining values as high as $H_0\sim 73-74$ km/s/Mpc might not be possible without producing internal inconsistencies, even in hybrid scenarios combining early and late-time departures from $\Lambda$CDM.

Wrapping up, our work aims first to quantify the evidence for dynamical dark energy within the CPL parametrization, using the most up-to-date dataset (which includes the SNIa sample from DES-Dovekie) and a robust methodology based on the FB approach. To begin with, we do that assuming standard physics before recombination. This study is particularly important in light of the recent discussions leading to divergent conclusions in Bayesian and frequentist analyses. Secondly, we reconstruct with the WFR method the most relevant background functions, improving the computation of the weights performed in our previous work \cite{Gonzalez-Fuentes:2025lei}. Finally, we perform a detailed study of how the evidence for dynamical DE and the status of the Hubble tension depend on possible deviations from $\Lambda$CDM prior to the decoupling time, following the approach outlined in the preceding paragraph.

The layout of this work is as follows. Section \ref{sec:method} is devoted to our methodology. In particular, we briefly remind the reader of the key aspects of the FB and WFR methods, provide practical details on how we speed up our computations using the Fisher matrix formalism, and describe the main components of our analysis pipeline.  In section \ref{sec:data}, we describe the datasets employed in the various parts of our study. In section \ref{sec:results}, we present and discuss our results, while section \ref{sec:conclusions} summarizes our conclusions.


\section{Methodology}\label{sec:method}

In this section, we start by briefly reviewing the main features of the Frequentist-Bayesian (FB) approach developed in Refs. \cite{Keeley:2021dmx,Amendola:2024prl} and also the Weighted Function Regression (WFR) method, which we employ to reconstruct the background cosmology from observational data, following our previous work \cite{Gonzalez-Fuentes:2025lei}. We place special emphasis on the computation of the WFR weights, which in this analysis is performed in a more rigorous and systematic manner using the FB approach. We explain how to optimize our pipeline to perform both the computation of the weights and the reconstruction at minimal computational cost, avoiding the use of Monte Carlo routines while maintaining full accuracy.

In our analysis, we use the Einstein–Boltzmann solver \texttt{CLASS}\footnote{\url{https://github.com/lesgourg/class_public}}
 \cite{Lesgourgues:2011re,Blas:2011rf} to compute theoretical predictions for the cosmological observables in the models considered. We employ \texttt{Cobaya}\footnote{\url{https://github.com/CobayaSampler/cobaya}}
 \cite{Torrado:2020dgo} to perform Monte Carlo Markov chain (MCMC) runs when required, primarily to test the validity of the Gaussian approximation. We also implement a Newton–Raphson minimization algorithm in \texttt{Python} to determine the minimum values of $\chi^2$ and the corresponding Fisher matrices (see Secs.~\ref{sec:FBmethod} and \ref{sec:distBF}). The \texttt{GetDist}\footnote{\url{https://github.com/cmbant/getdist}} package \cite{Lewis:2019xzd} is used for the analysis and visualization of parameter constraints from both Markov chains and the Fisher approximation. 
 
 Regardless of the model under consideration, in this work we assume a flat universe, with one massive neutrino with $m_\nu=0.06$ eV and two massless neutrinos, corresponding to the minimal normal-ordering scenario for neutrino masses, see, e.g., \cite{Esteban:2024eli}.


\subsection{The Frequentist-Bayesian method}\label{sec:FBmethod}

Bayes factors have become a widely used tool for model comparison in cosmology. However, as discussed in the Introduction, they suffer from important limitations due to their dependence on the choice of priors -- even when these are taken to be uninformative -- as well as from the arbitrariness of Jeffreys' scale. It is therefore necessary to explore alternative approaches that can mitigate these shortcomings.

The FB method \cite{Amendola:2024prl, Keeley:2021dmx} provides a viable alternative by promoting the evidence ratio to a frequentist statistic. This requires simulations and mock data to construct the distribution of evidence ratios under the assumption that one of the two models being compared is correct. The position of the observed evidence ratio within this distribution then determines a $p-$value, which can be used to reject or accept the hypothesis, largely independently of prior choices. 

In the following, we consider that the likelihood $\mathcal{L}$ is well approximated by a Gaussian in the parameters and show that the $p-$value obtained from the sampling distribution of the evidences does not depend on the priors $\pi_i$, when the latter are taken to be uninformative and flat (so they do not depend on the parameters themselves). In this case, the evidence reads

\begin{equation}
\mathcal{E}=\left[\prod_{i=1}^n\pi_i\right]\int d\vec{\theta}\mathcal{L} (D|\vec{\theta}) 
\end{equation}
in good approximation, and we can take advantage of the Fisher matrix formalism and expand the likelihood quadratically around the minimum as follows,

\begin{equation}
    \chi^2 = -2\ln \mathcal{L} = \chi^2_{\rm min} + (\theta_\alpha-\hat{\theta}_\alpha) F_{\alpha\beta} (\theta_\beta-\hat{\theta}_\beta) \, ,
\end{equation}
where we use Einstein's summation convention and $F_{\alpha\beta}$ are the components of the Fisher matrix, 

\begin{equation}
    F_{\alpha\beta} = -\left.\frac{\partial^2\ln \mathcal{L}}{\partial \theta_\alpha \partial \theta_\beta} \right|_{\hat{\theta}}=\frac{1}{2}\left.\frac{\partial^2\chi^2}{\partial \theta_\alpha \partial \theta_\beta} \right|_{\hat{\theta}} \, ,
\end{equation}
with $\hat{\theta}$ the parameter vector that maximizes the likelihood. Under these assumptions the likelihood can be integrated analytically, yielding

\begin{equation}
    \mathcal{E}  =(2\pi)^{n/2}\frac{e^{-\frac12 \chi^2_{\rm min}}}{\sqrt{|F|}}\left[\prod_{i=1}^n\pi_i\right] \, .
\end{equation}
Therefore, the Bayes ratio between a model $I$ nested to $\Lambda$CDM and the standard model takes the following form, 

\begin{equation}
    B=\frac{\mathcal{E}_I}{\mathcal{E}_{\Lambda {\rm CDM}}}=(2\pi)^{n_{\mathrm{extra},I}/2} e^{-\frac12\left( \chi^2_{{\rm min}, I}- \chi^2_{\rm min, \Lambda{\rm CDM}}\right)}\sqrt{\frac{|F_{\Lambda{\rm CDM}}|}{|F_I|}}\left[\prod_{i=1}^{n_{\mathrm{ extra},I}}\pi_i\right] \, ,
\end{equation}
where the priors for the parameters that are shared by the two models cancel and only those from the $n_{\mathrm{extra},I}$ parameters beyond $\Lambda$CDM -- belonging to model $I$ -- contribute. Equivalently, we can write the last expression simply as

\begin{equation}\label{eq:2lnB}
    2\ln(B) = n_{\mathrm{extra},I} \ln(2\pi)+ \left( \chi^2_{\rm min, \Lambda{\rm CDM}}-  \chi^2_{{\rm min}, I}\right) + \ln\left(\frac{|F_{\Lambda{\rm CDM}}|}{|F_I|}\right)+2\ln \mathcal{V}_{\mathrm{extra},I}^{-1} 
\end{equation}
after relating the uniform priors of the parameters to their corresponding prior volume, $\pi_i = \mathcal{V}^{-1}_i$, taking the same priors for those parameters that are shared by the two models and defining the total extra prior volume as $\mathcal{V}_{\mathrm{extra},I}=\Pi_{i=1}^{n_{\mathrm{extra},I}} \mathcal{V}_i$.

Given a dataset, we can perform the fitting analyses for models $I$ and the $\Lambda$CDM, compute the corresponding Fisher matrices, and calculate the Bayes ratio \eqref{eq:2lnB}. However, in order to know how significant is this value and how to translate it into quantitative statements on the relative performance of the models without making use of Jeffreys' -- or any alternative -- scale, we follow the FB method and construct the distribution of the quantity $2\ln(B)$. To do so, we generate mock data under the hypothesis that the best-fit $\Lambda$CDM extracted from the real data is the correct underlying model of the universe. It will be our fiducial model. The mock data is produced then from random Gaussian realizations centered at the best-fit $\Lambda$CDM values of the cosmological observables and using the real covariance matrix of the data. For each realization, we compute the Bayes ratio and save it to finally build the histogram (distribution), which will essentially tell us how likely is to obtain in a $\Lambda$CDM universe the Bayes factor obtained with the real data, translating that Bayes factor into a $p-$value.

From Eq.~\eqref{eq:2lnB} it becomes clear that, for uninformative uniform priors that are much broader
than the region where the likelihood is appreciable (so that the likelihood is not truncated within a
few $\sigma$), increasing the prior volume $\mathcal{V}_{\mathrm{extra},I}$ does not affect the resulting
$p-$value associated with the distribution of $B$ (or $2\ln B$). This is because the prior volume enters
Eq.~\eqref{eq:2lnB} only as an overall constant factor, which takes the same value when evaluating the
evidence ratio on either real or mock data.

It is also important to notice that in order to be sensitive to a $\sim 3\sigma$ effect -- like the one encountered for the evidence of dynamical DE in the context of the CPL parametrization  --, the number of mock datasets needed ascends to $\mathcal{O}(10^4)$, which might be computationally expensive even using the Fisher matrix formalism. Fortunately, in the following section we show that under the assumption of a Gaussian likelihood for the parameters -- which works very well for the models that we consider in this work -- the distribution of the Bayes factors follows an analytical $\chi^2$ distribution.


\subsection{Analytical distribution for the Bayes factor}\label{sec:distBF}

While the pipeline described in the previous section is formally sound, it may not be the most efficient approach. We recall that this pipeline requires the generation of mock data and, for each mock dataset, a numerical $\chi^2$  minimization and the computation of Fisher matrices for both models. The use of Fisher matrices -- accurate here because the posterior distribution of the parameters is very well approximated by a multivariate Gaussian -- allows us to avoid running Monte Carlo analyses, which would otherwise entail a prohibitive computational cost. However, despite the substantial speed-up provided by Fisher matrices, constructing the histogram of the quantity $2\ln B$ may still be computationally expensive, especially when an accurate characterization of the distribution tails is required, as in the case under study, due to the smallness of the $p-$values. Therefore, it would be highly desirable to derive the analytical distribution of $\ln B$, as this would allow us to compute automatically the $p-$value associated with the value of $\ln B$ inferred from the real data, having the shape of the tails fully under control. The derivation of such an analytical distribution is the main aim of this section. We follow the same approach applied in section III of \cite{Amendola:2024prl}, but use flat priors for the model parameters, instead of Gaussian priors. We also use a slightly different notation.

Let us consider two models, $M_A$ and $M_B$, with $M_B$ nested within $M_A$, and assume flat priors that are wide compared to the likelihood in such a way that the influence of the priors on the posterior distribution is, in practice, negligible. We aim to determine the distribution of $\ln B$ under realizations of the data
$d \sim \mathcal{N}(f^*, \Sigma)$, where $f^*$ is the vector of theoretical
predictions for the observables evaluated at a fiducial model $M_B$ -- namely, the
one that maximizes the posterior given the real data -- and $\Sigma$ is the
covariance matrix of the observed data.
For a given mock dataset $\mathcal{I}$, we can use Eq. \eqref{eq:2lnB} to obtain the value of $2\ln B^{(\mathcal{I})}$. It reads as follows,   

\begin{equation}\label{eq:2lnBI}
   2\ln B^{\I} = 2\ln \left(\frac{\mathcal{E}^{\I}_A}{\mathcal{E}^{\I}_B}\right) = (N_A-N_B)\ln (2\pi) + \chi_{B,{\rm min}}^{\I}-\chi_{A,{\rm min}}^{\I} + \ln \left(\frac{|F_B^{\I}|}{|F_A^{\I}|}\right) + 2\ln \mathcal{V}_{\mathrm{extra},I}^{-1}\,,
\end{equation}
where, to simplify the notation, we do $\chi_A^{\I}\equiv(\chi^{2}_A)^{\I}$. The $\chi^2$ for a given model and mock dataset takes the following form,

\begin{equation}\label{eq:chi2A}
    \chi^{\I}(\theta) = [d^{\I}_i-f_i(\theta)]\Sigma_{ij}^{-1}[d_j^{\I}-f_j(\theta)]\,,
\end{equation}
in terms of the elements $d_i^{\I}$  of the dataset $\mathcal{I}$ and the theoretical predictions of these observables in that concrete model, $f_i$, which of course depend on the model parameters $\theta$. Throughout this section, we adopt again Einstein's summation convention for repeated
indices and, for simplicity, we omit the subscript labeling the model,  which should be present in $\chi^{\I}$, the elements of $f$, as well as in its parameters $\theta$. 

If $f_i(\theta)$ are linear in the model parameters $\forall i$ in the region of parameter space where the likelihood is not negligible, we can express Eq. \eqref{eq:chi2A} as follows, 

\begin{equation}\label{eq:chi2Fisher}
    \chi^{\I} = \chi^{\I}_{\rm min}+(\theta_\alpha-\hat{\theta}^{\I}_\alpha)F^{\I}_{\alpha \beta}(\theta_\beta-\hat{\theta}_\beta^{\I})\,,
\end{equation}
with 

\begin{equation}\label{eq:FisherMatrix}
F^{\I}_{\alpha\beta}=\frac{\partial f_i}{\partial\theta_\alpha}\Sigma^{-1}_{ij}\frac{\partial f_j}{\partial\theta_\beta}\equiv g_{\alpha i}\Sigma_{ij}^{-1}g_{\beta j}
\end{equation}
the elements of the Fisher matrix and $\hat{\theta}^{\I}$ the vector of best-fit parameters obtained with the dataset $\mathcal{I}$. Latin and Greek indices run over the elements of the data vector and the parameter vector, respectively. Notice that the elements $g_{i\mu}\equiv \partial f_i/\partial\theta_\mu$ are constant, so $F^{\I}=F$ is independent of the data set $\mathcal{I}$. Thus, both the validity of the Fisher approximation and the constancy of the Fisher matrix across the mock datasets are explained by the linearity of $f(\theta)$.

Under this assumption, the theoretical predictions of the observables read

\begin{equation}\label{eq:flinear}
    f_i(\theta) = f_i^*+g_{i\gamma}(\theta_\gamma-\theta^*_\gamma)\,,
\end{equation}
where we use the linear relation around the fiducial parameter vector $\theta^*$, which satisfies $f^*=f(\theta^*)$. It is very convenient to expand around $\theta^*$ instead of $\hat{\theta}^{\I}$ because the former is of course constant over data realizations, whereas the best fit $\hat{\theta}^{\I}$ will be data-dependent. This is important, since we want to infer the distribution of $\chi^{\I}_{\rm min}=\chi^{\I}(\hat{\theta}^{\I})$ over Gaussian-distributed data, $d\sim \mathcal{N}(f^*,\Sigma)$. 

Now, we obtain the expression of $\chi^{\I}_{\rm min}$. To do so, we first employ Eq. \eqref{eq:flinear} evaluated at $\hat{\theta}^{\I}$, plug it into Eq. \eqref{eq:chi2A}, and finally use the relation 

\begin{equation}
\hat{\theta}^{\I}_\nu-\theta_\nu^*=F^{-1}_{\mu\nu}g_{i\mu}\Sigma^{-1}_{ij}(d_j^{\I}-f_{j}^*)\,,
\end{equation}
which is obtained from equating the expressions of $\partial\chi^{\I}/\partial\theta_\nu$ inferred from Eqs. \eqref{eq:chi2A} and \eqref{eq:chi2Fisher}. After a bit of algebra and making use of Eq. \eqref{eq:FisherMatrix} we are led to the following result, 

\begin{equation}\label{eq:chi2min}
    \chi^{\I}_{\rm min}=(d^{\I}-f^*)^T M(d^{\I}-f^*)\,,
\end{equation}
with $M=\Sigma^{-1}-\Sigma^{-1}gF^{-1}g^T\Sigma^{-1}$ independent of $\mathcal{I}$. Eq. \eqref{eq:chi2min} shows that $\chi_{\rm min}^{(\mathcal{I})}$ is a quadratic form of Gaussian random variables (the data $d^{\I}$) with 0 mean, i.e., $\langle d^{(\mathcal{I})}-f^*\rangle=0$. Therefore, we can define $z\equiv \chi_{B,{\rm min}}^{\I}-\chi_{A,{\rm min}}^{\I}=2\ln B^{\I} -a $, with the constant factor (cf. Eq. \ref{eq:2lnBI}) 

\begin{equation}
a\equiv (N_A-N_B)\ln(2\pi) +\ln \left(\frac{|F_B|}{|F_A|}\right)+2\ln \mathcal{V}_{\mathrm{extra}, I}^{-1}\,,
\end{equation}
and $z$ is distributed as a standard $\chi^2$ distribution with $\nu=N_A-N_B$ degrees of freedom (see proof below),

\begin{equation}\label{eq:analyticaldist}
    P(z;\nu)=\frac{1}{2^{\nu/2}\Gamma(\nu/2)}z^{\nu/2-1}e^{-z/2}\,.
\end{equation}
We remark here that the analytical distribution \eqref{eq:analyticaldist} is only valid if the assumption of the linearity of $f(\theta)$ in the region of interest of the parameter space for the two models involved in the analysis is fulfilled in good approximation.

As a simple exercise, we can check, for instance, that we retrieve the expected value $\langle\chi^{\I}_{\rm min,B}-\chi^{\I}_{\rm min,A}\rangle$ for this distribution. To do so, it is convenient to rewrite the matrix $M$ as

\begin{equation}
    M=\Sigma^{-1}(I-gF^{-1}g^T\Sigma^{-1}) \equiv \Sigma^{-1}(I-C)\,,
\end{equation}
where $I$ is the identity matrix of dimension equal to the number of data points, $N$. Now, we can employ these results to compute the mean $\chi^{\I}_{\rm min}$ over the data realizations:

\begin{equation}
\langle\chi^{\I}_{\rm min}\rangle= \langle(d^{\I}-f^*)^T M(d^{\I}-f^*)\rangle={\rm Tr}(M\Sigma)={\rm Tr}(I-C)=N-{\rm Tr}(C)\,.
\end{equation}
Here, we have used again the fact that, by definition, $\langle d^{\I}\rangle=f^*$. Using 

\begin{equation}
{\rm Tr}(C)= {\rm Tr}(gF^{-1}g^T\Sigma^{-1})={\rm Tr}(F^{-1}F)=N_M\,,
\end{equation}
with $N_M$ the number of parameters in the model, we finally obtain, 

\begin{equation}
\langle\chi^{\I}_{\rm min}\rangle=N-N_M\quad\rightarrow\quad \nu=\langle\chi^{\I}_{\rm min,B}-\chi^{\I}_{\rm min,A}\rangle=N_A-N_B\,.
\end{equation}
As a direct consequence of these results, we find that the $p-$value associated with the value of $2\ln B$ -- with
$B=\mathcal{E}_I/\mathcal{E}_{\Lambda{\rm CDM}}$ --inferred from the real data
coincides with the one used to compute the exclusion level of $\Lambda$CDM
relative to model~$I$ with a likelihood-ratio test, as it is done, e.g., in \cite{DESI:2025zgx,DESI:2025fii,Gonzalez-Fuentes:2025lei}. This equivalence stems from the validity of the Fisher approximation and Wilks' theorem \cite{Wilks:1938dza}. Hence, in the limit of non-informative priors, the statistical significance obtained directly from the maximum of the posterior distribution, i.e., from the corresponding value of $\Delta \chi^2_\mathrm{MAP}$ \cite{DESI:2025zgx}, agrees with this result.


\subsection{The Weighted Function Regression method}

In this section, we briefly summarize the WFR method \cite{Gomez-Valent:2018hwc,Gomez-Valent:2018gvm} (see also \cite{Liddle:2006kn}) which we recently applied to the reconstruction of the equation-of-state of DE and cosmographical functions in \cite{Gonzalez-Fuentes:2025lei}. An interpretation of dark energy dynamics through the fitting results of a specific parameterization may lead to biased results. Although one could achieve greater flexibility by incorporating more parameters, there is a risk of overfitting and adding unnecessary complexity. The WFR naturally applies Occam's razor by assigning Bayesian weights to different parameterizations that form a basis for the reconstruction of cosmological observables. 

As described in the aforementioned references, the probability density associated with a certain shape of a function $f(z)$ given some data $\mathcal{D}$ is 
\begin{equation}\label{eq:oldreco}
P[f(z)|\mathcal{D}]=\frac{\sum\limits_{J=0}^{n}P(f(z)|M_J)B_{J*}}{\sum\limits_{J=0}^{n}B_{J*}}\,,
\end{equation}
where $M_J$ are the models considered as a basis for the reconstruction, each with $n_J$ additional parameters. In \cite{Gonzalez-Fuentes:2025lei}, we used expansions in powers of $(1-a)$ in both  $w_{\rm DE}(a)$ (e.g. $w$CDM, CPL, ... ) and in  $\rho_{\rm DE}(a)$ -- but separately, since given a form of the  EoS parameter one can automatically get the associated energy density, and viceversa, if DE is self-conserved. From this expression, the weights assigned to each of the models arise from Bayes factors $B_{J*}$ computed with respect to a reference model $M_*$:
\begin{equation}\label{eq:weightold}
W_J = \frac{B_{J*}}{\sum\limits_{I=0}^{n}B_{I*}}\,.
\end{equation}
In practice,  we approximated the ratio of evidences using either the Akaike Information Criterion (AIC) \cite{Akaike} or the Deviance Information Criterion (DIC) \cite{DIC}:
\begin{equation}\label{eq:weightIC}
    B_{J*} \simeq e^{\Delta {\rm IC}_J/2}\,,
\end{equation}
where $\Delta {\rm IC}_J= {\rm IC}_*-{\rm IC}_J$. In Appendix C of \cite{Gonzalez-Fuentes:2025lei}, we compared the performance of AIC and DIC, and also discussed the results obtained with the Bayesian Information Criterion (BIC) \cite{BIC}, which was not suitable in this case due to an unfair penalization caused by the large number of data points employed in the analysis. Using these information criteria instead of exact Bayesian evidences reduced the computational cost of our analysis and avoided the intricate discussion of prior widths. In this context, there is scant theoretical guidance to construct meaningful priors\footnote{For instance, in the CPL model one may impose the condition $w_0+w_a<0$ to ensure matter domination
in the past. However, a significant degree of subjectivity remains in the specific choice of priors.}. On the other hand, choosing priors that are much broader than the likelihood would
unfairly penalize models with additional parameters if Bayes factors were employed. In the following, we discuss these matters with further detail by incorporating the FB method to the computation of weights for WFR reconstruction.

As seen in section \ref{sec:FBmethod}, the main outcome of the FB method is a $p-$value, which essentially tells us what is the probability of having a Bayes ratio $B=\mathcal{E}_I/\mathcal{E}_{\Lambda{\rm CDM}}$ equal or larger than the Bayes ratio $B_r$ extracted from the real data, if we assume the benchmark model, i.e., the $\Lambda$CDM, to be the correct model of the universe. In other words, the FB method gives us the probability $(p{\rm -value})=P(B\geq B_r|\Lambda{\rm CDM\,correct})$. Using Bayes theorem, we find,
\begin{equation}\label{eq:BT1}
P(B\geq B_r|\Lambda{\rm CDM\,correct})P(\Lambda{\rm CDM\,correct})=P(\Lambda{\rm CDM\,correct}|B\geq B_r)P(B\geq B_r)\,,
\end{equation}
and the same for the model $M_I$ that is being confronted with the $\Lambda$CDM,

\begin{equation}\label{eq:BT2}
P(B\geq B_r|M_I{\rm \,correct})P(M_I{\rm \,correct})=P(M_I{\rm\,correct}|B\geq B_r)P(B\geq B_r)\,.
\end{equation}
The weight $W_I$ associated to model $M_I$, relative to the $\Lambda$CDM, is just proportional to

\begin{equation}
W_I\propto \widetilde{W}_I=\frac{P(M_I{\rm\,correct}|B\geq B_r)}{P(\Lambda{\rm CDM\,correct}|B\geq B_r)}\,.
\end{equation}
Considering that we have no prior preference for any of the two models, i.e., that $P(\Lambda{\rm CDM\,correct})=P(M_I{\rm \,correct})$, we can reexpress $\widetilde{W}_I$ as follows, using the ratio of Eqs. \eqref{eq:BT1}-\eqref{eq:BT2},

\begin{equation}
    \widetilde{W}_I=\frac{P(B\geq B_r|M_I{\rm \,correct})}{P(B\geq B_r|\Lambda{\rm CDM\,correct})}=\frac{1-(p{\rm -value})}{(p{\rm -value})}\,.
\end{equation}
This expression is in accordance with the expectation that models with lower $p-$value should have higher weights. In particular, notice that in the case $p=0.5$, there is no preference for neither $M_I$ nor $\Lambda$CDM and hence the weight is indecisive $\widetilde{W}_I=1$, as desired. 

The probability of having a specific shape of $f(z)$, given that we obtain a set of $p-$values for a number of models is
\begin{equation}
P[f(z)|\{B_{r,I}\} ]\propto\sum_{J=0}^{n}P(f(z)|M_J)\widetilde{W}_J\,.
\end{equation}
Integrating both sides over the measure, we find the factor that normalizes the weights,
\begin{equation}\label{eq:weights}
   W_J = \frac{\widetilde{W}_J}{\sum\limits_{I=0}^{n} \widetilde{W}_I}\,,
\end{equation}
so we can finally write
\begin{equation}
P[f(z)]=\sum_{J=0}^{n}P(f(z)|M_J)W_J\,.
\end{equation}
If the function we want to reconstruct is linear in very good approximation within the volume of parameter space where the posterior distribution is non-negligible, we can express the aforesaid function as follows,

\begin{equation}
f(\theta) = f(\hat{\theta})+\frac{\partial f}{\partial\theta_i}\Big|_{\theta=\hat{\theta}}(\theta_i-\hat{\theta}_i)\,,
\end{equation}
and $f$ distributes as a Gaussian centered at $\hat{f}=f(\hat{\theta})$ and with standard deviation 

\begin{equation}
\sigma[f]\equiv \langle[f(\theta)-\hat{f}]^2\rangle ^{1/2} =\left[\frac{\partial f}{\partial\theta_i}\Big|_{\theta=\hat{\theta}}\frac{\partial f}{\partial\theta_j}\Big|_{\theta=\hat{\theta}}(F^{-1})_{ij}\right]^{1/2}\,.
\end{equation}
If this happens for all the models $M_J$ entering the WFR reconstruction, we can write the weighted probability distribution of $f$ as a Gaussian Mixture Model, i.e.,  

\begin{equation}
P[f] = \frac{1}{\sqrt{2\pi}}\sum_{J=0}^n \frac{W_J}{\sigma_J}e^{-\frac{1}{2}\left(\frac{f-\hat{f}_J}{\sigma_J}\right)^2}\,.
\end{equation}


\section{Cosmological data}\label{sec:data}

In those parts of the analysis in which we assume standard pre-recombination physics, i.e., in Secs. \ref{sec:FBresults} and \ref{sec:standardphysics}, we employ the following triad of state-of-the-art datasets:

\begin{itemize}
    \item {\it Cosmic Microwave Background}: We use compressed CMB information through the correlated multivariate Gaussian on $(\theta_*, \omega_{\rm b}, \omega_{\rm bc})_{\rm CMB}$ \cite{Lemos:2023xhs}, as in \cite{DESI:2025zgx}, with $\omega_{\rm bc}$ the sum of the cold dark matter (CDM) and baryon reduced density parameters. This compressed likelihood is extracted from the {\it Planck} \texttt{CamSpec} CMB data \cite{Efstathiou:2019mdh,Rosenberg:2022sdy}. It encapsulates the most relevant CMB constraints at the background level in models that leave the physics before recombination untouched. 
    
    \item {\it Baryon Acoustic Oscillations}: We use BAO data from DESI Data Release 2 (DR2) as described in Table IV of Ref. \cite{DESI:2025zgx}, which includes measurements of
    
    \begin{equation}
    \frac{D_V(z)}{r_d}= \frac{1}{r_d}\left[zD_M^2(z)D_{H}(z)\right]^{1/3}\,,\quad \frac{D_M(z)}{r_d}\,,\quad \frac{D_H(z)}{r_d} \,,
    \end{equation}
    across redshift bins from $z=0.295$ to $z=2.33$, duly accounting for the existing correlations, with $D_M(z)=(1+z)D_A(z)$ and $D_H(z)=c/H(z)$ the comoving angular diameter and Hubble distances, respectively, and $D_V(z)$ the so-called volume-averaged dilation scale.
    
    \item {\it Type Ia Supernovae}: We make use of the Pantheon+ and DES-Dovekie SNIa compilations. Pantheon+ \cite{Scolnic:2021amr} is based on 1550 spectroscopically confirmed SNIa in the redshift range 
$0.01<z<2.26$, while DES-Dovekie \cite{DES:2025sig} consists of a recalibrated sample of 1820 SNIa spanning 
$0.025<z<1.14$. This compilation addresses calibration issues that affected the previous DES-Y5 sample \cite{DES:2024hip, DES:2024jxu}. The publicly available DES-Dovekie data only permits analyses in which the supernova absolute magnitude $M$ is marginalized over. To ensure a consistent treatment, in Secs. \ref{sec:FBresults} and \ref{sec:standardphysics} we also marginalize over $M$.
\end{itemize}

In section \ref{sec:newphysics}, where we do not assume standard pre-recombination physics, we employ CMB/BBN information in a more compressed form, in terms of the following Gaussian priors on the CMB acoustic scale and the baryon energy density: $100\theta_* = 1.04223 \pm 0.00055$ \cite{Sharma:2025iux} and $\omega_b=0.02196\pm 0.00063$ \cite{Schoneberg:2024ifp}. The former is extracted from an analysis of {\it Planck} baseline TT,TE,EE likelihoods assuming a $\Lambda$CDM+$N_{\rm eff}$ cosmology
and one massive neutrino with $m_\nu=0.06$ eV, whereas the latter has been obtained from a BBN analysis also allowing for a non-standard number of ultra-relativistic species in the cosmic plasma. The corresponding uncertainties are about $50\%$ and $10\%$ larger than the one obtained assuming only the contribution of three neutrino species. We deem them to be a more conservative option, more in accordance with the model-agnostic approach followed in this study. We do not consider priors on the reduced CDM density, since different early-time solutions to the Hubble tension might require different amounts of dark matter around the last-scattering surface and before the decoupling time, so preventing its use will make our results more general.  Moreover, given that in this part of the analysis we focus on the $H_0$ tension, which can ultimately be phrased as a tension in the calibrators of the distance ladders (see, e.g., \cite{Favale:2023lnp,Favale:2025mgk}), we employ PantheonPlus+SH0ES to account for the Cepheid calibration of the SNIa absolute magnitude and allow $M$ to vary freely in the fitting analysis, as first advocated in \cite{Camarena:2021jlr,Efstathiou:2021ocp,Benevento:2020fev}. For the sake of completeness, we present the results also by replacing SH0ES calibration by a prior on $H_0$ extracted from the Tip of the Red Giant Branch (TRGB) calibration, which is a much more conservative option. We employ $H_0^{\rm CCHP}=70.39\pm 1.94\,$km/s/Mpc, from the Chicago-Carnegie Hubble Program (CCHP) \cite{Freedman:2024eph}\footnote{Alternative calibration methods independent from distance ladder and CMB measurements can be achieved through the use of cosmic chronometers \cite{Gomez-Valent:2018hwc,Gomez-Valent:2021hda,Favale:2023lnp,Favale:2025mgk} or the method applied in \cite{Zaborowski:2025umc}. However, current data on cosmic chronometers still yield large uncertainties on $H_0$, and the method of \cite{Zaborowski:2025umc} would introduce undesired correlations with the other datasets employed in our analysis.}.


\section{Results}\label{sec:results}

\subsection{Evidence for dynamical DE within CPL in the light of the FB method}\label{sec:FBresults}

\begin{table}[t!]
    \centering
    \begin{tabular}{c|cc}
      & CMB$+$DESI DR2$+$PantheonPlus & CMB$+$DESI DR2$+$DES Dovekie \\ \hline 
     $\chi^2_{\mathrm{min}} (\Lambda {\rm CDM})$ & $1422.69$ & $1652.03$  \\
     $\chi^2_{\mathrm{min}} (\mathrm{CPL})$  & $1414.43$ & $1640.41$\\
     $\Delta$AIC & $+4.26$ & $+7.62$  \\
     $\Delta$DIC & $+5.37$ & $+7.70$ \\
     $\Delta$BIC & $-6.50$ & $-3.41$  \\
    \end{tabular}
    \caption{Comparison of the maximum likelihood results for both SNIa datasets, displaying also the difference of information criteria defined as $\Delta{\rm IC}={\rm IC}_{\Lambda{\rm CDM}}-{\rm IC}_{\rm CPL}$.}
    \label{tab:CPLresults}
\end{table}

In this section, we apply the FB method described in section \ref{sec:FBmethod} to the CPL parametrization in order to assess its preference over $\Lambda$CDM. Our analysis uses a combination of compressed CMB constraints $(\theta_*, \omega_{\rm b}, \omega_{\rm bc})$, DESI DR2 BAO measurements, and the new DES-Dovekie SNIa dataset. We refer the reader to Ref. \cite{Sakr:2025chr} for a previous application of this methodology comparing $w$CDM and CPL, based on the Pantheon+ SNIa sample.

\begin{figure}[t!]
    \centering
    \includegraphics[width=0.8\linewidth]{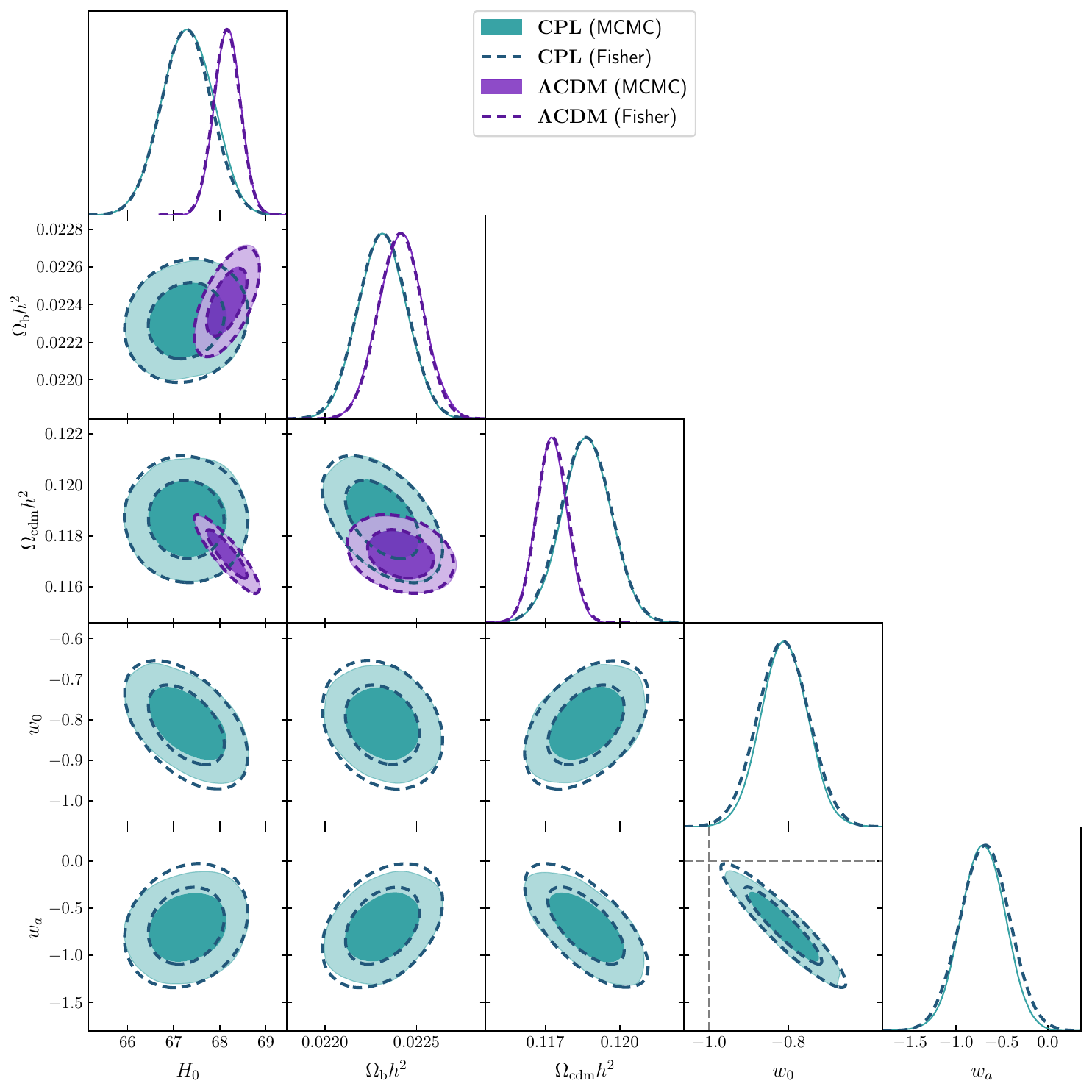}
    \caption{Comparison of the contour plots from MCMC (solid) and Fisher approximation (dashed) with data from CMB, DESI DR2 BAO and DES-Dovekie SNIa for both the $\Lambda$CDM and CPL models. The posterior distribution is Gaussian in very good approximation. See the main text for the corresponding discussion.}
    \label{fig:MCMCFisher}
\end{figure}

As a preliminary step, we have checked that for this dataset and model the Fisher approximation works extremely well. In Fig. \ref{fig:MCMCFisher}, we compare the posteriors extracted from the MCMC analyses for both $\Lambda$CDM and CPL (using flat wide priors, i.e., much wider than the likelihood) with those inferred from minimization and quadratic expansion of $\ln \mathcal{L}$ (Fisher). The contours show that the Fisher approximation is valid in this case. Hence, the assumptions used in section \ref{sec:distBF} are fulfilled and we can obtain the $p-$value associated to the computed value of the Bayes factor from the analytical distribution given in Eq.  \eqref{eq:analyticaldist}. From this figure, it becomes clear that the ability of the CPL parametrization to provide a better fit to these data does not arise solely from the additional freedom in the dark-sector dynamics encoded in the $w_0$–$w_a$ parametrization. The broadening of the posterior in the extended model is accompanied by shifts in the central values of the parameters $H_0$, $\Omega_{\rm b}h^2$ and $\Omega_{\rm cdm} h^2$. This compensation is necessary to maintain the distance to the last-scattering surface stable once late-time deviations are allowed. On top of that, the CPL takes advantage of the degeneracies between parameters and is able to provide a better fit to a combination of datasets that is in tension within the standard model. 

\begin{figure}
    \centering
    \includegraphics[width=0.8\linewidth]{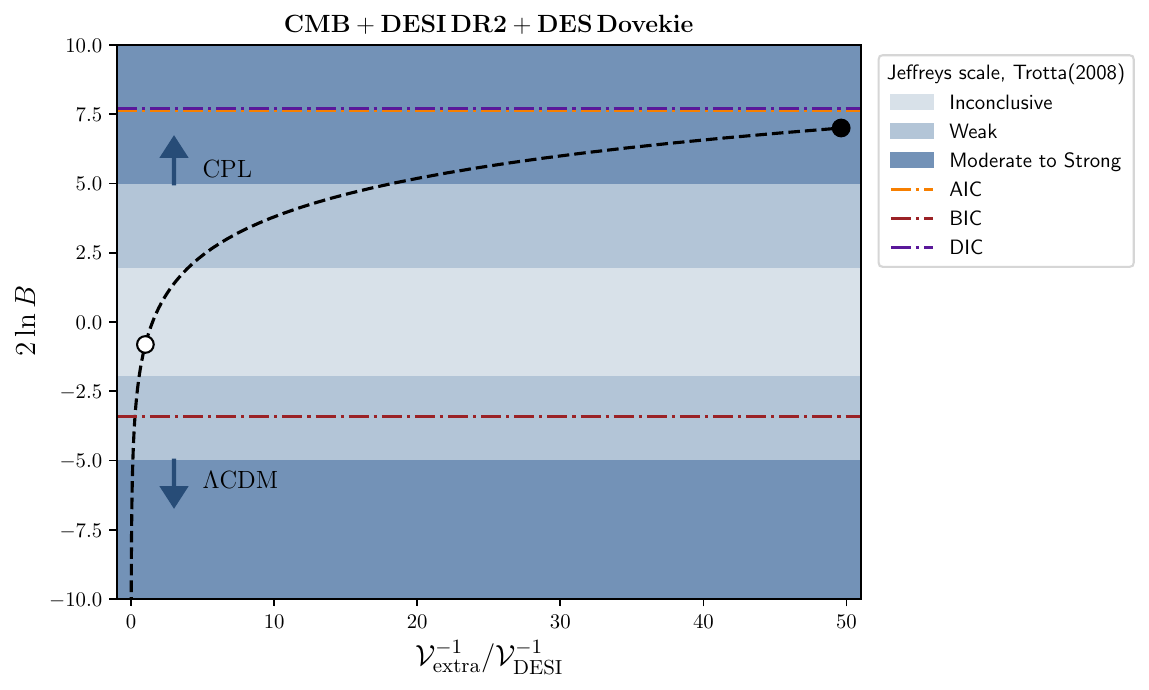}
    \caption{Logarithm of the Bayes factor with data from CMB + DESI DR2 + DES Dovekie comparing $\Lambda$CDM and CPL as a function of the extra prior volume in parameter space. This is defined for uniform priors as $\pi_{i}=\mathcal{V}_{{\rm extra},i}^{-1}$ for each parameter beyond $\Lambda$CDM and the total $\mathcal{V}_{\rm extra}=\Pi_{i=1}^{n_\mathrm{extra}}\mathcal{V}_{{\rm extra},i}$. For reference, this has been normalized to the priors used by DESI DR2: $w_0\sim \mathcal{U}[-3,1]$, $w_a\sim \mathcal{U}[-3,2]$, without imposing $w_0+w_a<0$. We also display the Jeffreys' scale \cite{Trotta:2008qt} and information criteria AIC, DIC, BIC. The black point on the right corresponds to the result obtained using the prior constructed from the marginalized $3\sigma$ limits of the extra parameters, whereas the white point on the left is obtained using the DESI DR2 prior.}
    \label{fig:logBJeffreys}
\end{figure}

In Table \ref{tab:CPLresults}, we present a summary of the minimization results. With $\Lambda$CDM the best fit has $\chi^2_{\rm min}=1652.03 $ while CPL decreases it by 12 units, $\chi^2_{\rm min}= 1640.41$. We also display the values of the different information criteria, showing that AIC and DIC agree, pointing both to a moderate preference for the CPL according to Jeffreys' scale (cf. Table \ref{tab:jeffreysscale}), whereas BIC is extremely conservative in this case. This is expected for a dataset including $n_d=1836$ points, which implies that the penalization factor for each extra parameter is $\ln(n_d)\sim 7.5$. For completeness, the results for Pantheon+ are also reported. DES-Dovekie enhances slightly the dynamical DE signal compared to that obtained with Pantheon+. This is reflected into both larger differences of $\chi^2_{\rm min}$ and information criteria.

From Eq. \eqref{eq:2lnB}, one can visualize how the Bayes factors change as a function of the prior volume, as done in Fig 1 of \cite{Patel:2024odo}. In Fig. \ref{fig:logBJeffreys} we plot the evolution of $2\ln B$ for CMB$+$DESI DR2$+$DES Dovekie as a function of $\mathcal{V}_{\rm extra}$ defined for the extra parameters $w_0$ and $w_a$, which control the DE dynamics. Taking uniform priors, for each parameter one has $\pi_i = \mathcal{V}_{\mathrm{extra},i}^{-1}$ and its product defines the total volume. These volumes have been normalized to those used in DESI DR2 main analysis (i.e., $w_0\sim \mathcal{U}[-3,1]$ and $w_a\sim \mathcal{U}[-3,2]$) to ease the interpretation. The color gradient displays the evidence according to Jeffreys' scale (cf. again Table \ref{tab:jeffreysscale}). This figure exposes clearly how this scale becomes arbitrary in this context. One can shift from moderate evidence to weak, or inconclusive by just choosing different priors that play no role in the inference process, since they are very broad and hence leave the shape of the posterior unaffected. If one takes a large prior volume (left part of the plot), then the Jeffreys' scale would concede moderate evidence in favor to $\Lambda$CDM, as opposed to the likelihood-ratio test, AIC and DIC. As we consider tighter priors and move to the right part of the figure, one can reach the opposite and obtain moderate evidence in favor of CPL. 

\begin{figure}
    \centering
    \includegraphics[width=0.7\linewidth]{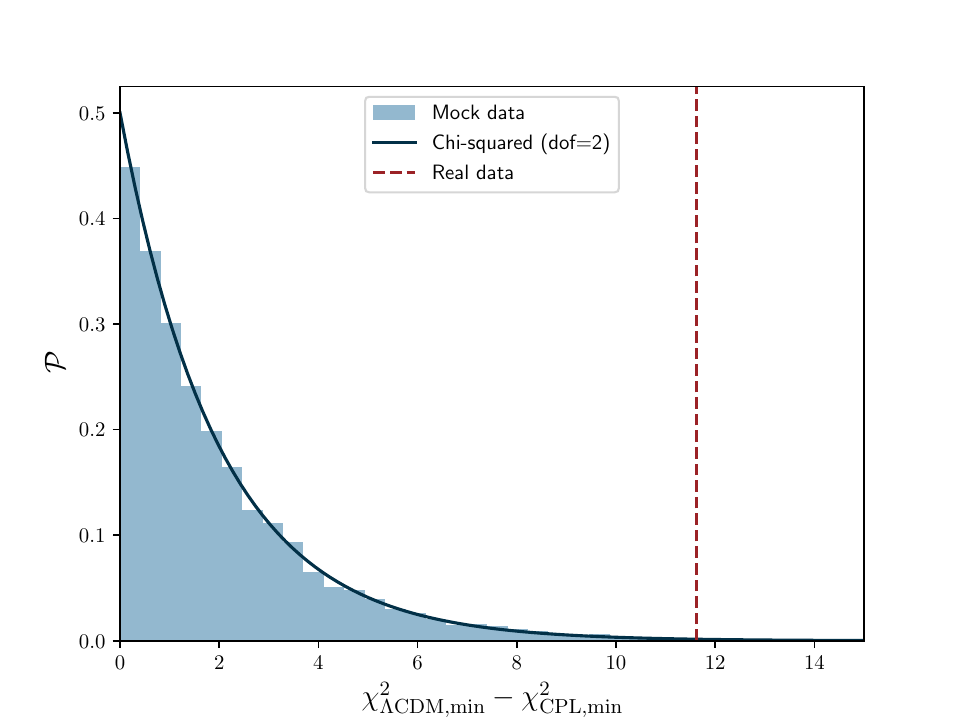}
    \caption{Distribution of the difference of $\chi^2_{\rm min}$ obtained from $N\sim 6000$ mock realizations assuming the validity of $\Lambda$CDM, and the corresponding theoretical curve, described by a $\chi^2$ with 2 degrees of freedom, cf. section \ref{sec:distBF}. The red vertical dashed line corresponds to the value obtained with real data, with an associated $p-$value=0.00299. Therefore, the $\Lambda$CDM is excluded at 99.70\% CL in front of the CPL, i.e., at $2.97\sigma$ CL.}
    \label{fig:histogram}
\end{figure}

Using the FB method, we can eliminate the dependence of our conclusions on the uninformative priors adopted in the analysis, avoid the need for Jeffreys' scale, and facilitate the interpretation of the results. In Fig. \ref{fig:histogram} we show the distribution (histogram) of the difference of $\chi^2_{\rm min}$ for the $\Lambda$CDM and the CPL obtained following the method described in section \ref{sec:FBmethod}, using mock data generated under the assumption of the validity of $\Lambda$CDM, with the SNIa from DES-Dovekie in combination with the data from DESI DR2 and CMB. This distribution is essentially the same as $2\ln B$, up to a constant shift that depends on the ratio of the Fisher matrix determinants. We also display the analytical $\chi^2$ distribution with two degrees of freedom (see Eq.~\eqref{eq:analyticaldist}), which is in excellent agreement with the histogram, as expected given the aforesaid Gaussian features of the posterior distributions. According to the FB method, $\Lambda$CDM is excluded at 99.70\% ($2.97\sigma$) CL relative to the CPL parametrization. This quantification of the tension matches pretty well with the significance inferred directly from the contour plot in the $(w_0,w_a)$-plane (see Fig. \ref{fig:MCMCFisher}). When using Pantheon+, as previously mentioned, the exclusion becomes slightly weaker, at 98.39\% ($2.41\sigma$) CL. These values resonate much better with the moderate evidence inferred from AIC and DIC using Jeffreys' scale, rather than the one inferred from the exact Bayes ratio using the DESI priors, which weakly favors the $\Lambda$CDM (see again Fig. \ref{fig:logBJeffreys}), as reported also in \cite{Ong:2025utx,Hergt:2026moc,Ong:2026tta}. In the following section, we will show that, in fact, the FB method leads to similar WFR weights to those obtained using AIC and DIC.

Although some works attribute the discrepancy between frequentist and Bayesian quantifications of the evidence in favour of CPL to the Jeffreys-Lindley paradox \cite{Ong:2026tta}, the latter is just a manifestation of the Bayesian evidence dependence on the priors. This can be considered an inherent feature of the Bayesian approach, which need not be a problem with well-motivated priors. In fact, in some cases the Bayes factor can correctly favor models that would be rejected with a frequentist methodology (cf. Appendix A of \cite{Trotta:2005ar}). Nevertheless, in the specific case of $(w_0,w_a)$, there is no well-motivated choice of priors and hence the Occam penalty term becomes arbitrary. In addition, choosing them wider than the likelihood does not settle the debate: as shown in Fig. \ref{fig:logBJeffreys}, the penalty term changes $\ln B$ by factors of a few, which is sufficient to shift from a conclusion to the opposite when interpreted as betting odds.

This preliminary analysis underscores the importance of adopting a robust statistical framework to evaluate the evidence for physics beyond the standard cosmological model, free from subjective assumptions and associated ambiguities. The FB method provides a reliable and well-founded approach for carrying out this type of calculation in a statistically consistent manner.


\subsection{DE reconstruction with standard pre-recombination physics}\label{sec:standardphysics}

In this section, we apply the WFR technique to reconstruct the background DE properties and the expansion history of the universe. The weights defined in Eq.~\eqref{eq:weights} are computed using the FB method. We adopt a basis for the WFR reconstruction based either  on truncated expansions of the EoS parameter or the DE density as a function of the scale factor -- which we termed the $w$-basis and $\rho$-basis, respectively. As explained above, the Fisher matrix formalism and the analytical expression for the distribution of $\ln(B)$ can be reliably employed only when the posterior distributions of both $\Lambda$CDM and the dynamical DE model under consideration are close to Gaussian. In our previous work \cite{Gonzalez-Fuentes:2025lei}, we showed that, within the $\rho$-basis the posterior distributions of the parameters are very well approximated by Gaussians for those models that contribute non-negligibly to the reconstruction of the cosmological functions. This contrasts with the behavior found in the $w$-basis, where quartic expansions of the EoS parameter in the scale factor (the so-called  CPL$^{++}$ parametrization \cite{Zhang:2017idq,Dai:2018zwv,Nesseris:2025lke}) lead to significantly non-Gaussian posteriors (see the contour plots in Appendix B of \cite{Gonzalez-Fuentes:2025lei}). Therefore, the $\rho$-basis provides a more suitable framework for the WFR reconstruction when models with $n_J>3$ are considered, whereas the $w$-basis can be still safely used if it is sufficient to restrict the analysis to $n_J\leq 3$. 
\begin{table*}
\begin{center}
\begin{tabular}{|c|ccc|c|}
\multicolumn{5}{c}{\textbf{CMB + DESI DR2 + DES Dovekie, $w-$basis}}\\ \hline
Parameter & $w$CDM & CPL & CPL$+$ & Reconstructed\\ \hline
$H_0$[km/s/Mpc] &$67.65\pm0.55$  & $67.27\pm0.55$ & $67.42\pm0.59$ & $67.32\pm 0.56$\\
$10^2\omega_{\rm b}$ & $2.246\pm0.013$ & $2.231\pm0.013$ & $2.230\pm 0.014$ & $2.231\pm 0.013$\\
$10\omega_{\rm cdm}$ & $1.167\pm 0.008$ & $1.186\pm0.010$ & $1.189\pm0.010$ & $1.187\pm0.010$\\
$w_0$ & $-0.98\pm 0.02$ & $-0.81\pm0.06$ & $-0.89\pm 0.12$ & $-$\\
$w_a$ & $-$ & $-0.68\pm0.27$ & $0.01\pm1.04$ & $-$\\
$w_b$ & $-$ & $-$ & $-1.19\pm 1.79$& $-$\\ \hline
$\Omega_{\rm m}^0$ &$0.305\pm 0.005$ & $0.313\pm 0.006$ &$0.312\pm 0.006$ &$0.313\pm0.006$\\ 
$F_\mathrm{cross}\left[\%\right]$ & $0$ & $99.07$ & $99.30$ & $98.46$\\
$\chi^2_{\rm min}$ & $1650.64$ & $1640.41$ & $1640.01$ & $-$\\
$p-$value & $0.237$ & $0.00299$ & $0.00730$ & $-$\\
$W_J\left[\%\right]$ & $0.68$ & $70.53$ & $28.79$ & $-$\\ \hline
\end{tabular}
\caption{Mean values and uncertainties at 68\% CL obtained with CMB $+$ DESI DR2 $+$ DES Dovekie for the three models in the $w$-basis employed in the WFR. In the last column, we display the reconstructed constraints. We report also the probabilities of phantom crossing (from phantom in the past to quintessence at lower redshifts), $F_{\rm cross}$, the minimum $\chi^2$, $\chi^2_{\rm min}$, as well as the corresponding $p-$values and weights $W_J$, cf. section \ref{sec:method} for details.}\label{tab:tab_des_w}
\end{center}
\end{table*}

\begin{table*}
\begin{center}
\begin{tabular}{|c|ccc|c|}
\multicolumn{5}{c}{\textbf{CMB + DESI DR2 + DES Dovekie, $\rho-$basis}}\\ \hline
Parameter & $n_J=1$ & $n_J=2$ & $n_J=3$ & Reconstructed  \\ \hline
$H_0$[km/s/Mpc] & $67.34\pm0.55$ & $67.16\pm 0.56$ & $67.45\pm0.61$ & $67.27\pm 0.59$\\
$10^2\omega_{\rm b}$ & $2.247\pm0.012$ & $2.233\pm 0.013$ & $2.223\pm0.014$ & $2.232\pm 0.014$\\
$10\omega_{\rm cdm}$ & $1.165\pm0.008$ & $1.184\pm 0.010$ & $1.190\pm0.011$ & $1.186\pm 0.011$\\
$C_1$ & $0.158\pm 0.095$ & $0.75\pm 0.24$ & $0.30\pm0.49$ & $-$\\
$C_2$ & $-$ & $-1.54\pm 0.56$ & $0.85\pm2.32$ & $-$\\
$C_3$ & $-$ & $-$ & $-2.97\pm 2.76$ & $-$\\ \hline
$\Omega_{\rm m}^0$ & $0.308\pm 0.005$ & $0.313\pm 0.006$ & $0.312\pm0.006$& $0.313\pm 0.006$\\
$F_\mathrm{cross}\left[\%\right]$ & $0.00$ & $99.50$ & $99.83$ & $96.87$\\
$F_\mathrm{neg}\left[\%\right]$ & $0.00$ & $0.00$ & $4.67$ & $1.64$\\
$\chi^2_{\rm min}$ & $1649.13$  & $ 1641.13$ & $1640.09$ & $-$\\
$p-$value & $0.0884$ & $0.00429$ & $ 0.00758$ & $-$\\
$W_J\left[\%\right]$ & $2.76$  & $62.17$&  $35.06$ & $-$\\ \hline
\end{tabular}
\caption{Same as in Table \ref{tab:tab_des_w}, but using the $\rho$-basis.}\label{tab:tab_des}
\end{center}
\end{table*}

Each model $M_J$ of the $w$-basis is linked to a particular form of the EoS parameter, 

\begin{equation}\label{eq:CPL_gener}
   w_{\mathrm{DE},J}(a) = \sum_{l=0}^{J}w_l (1-a)^l\,,
\end{equation}
with the following energy density, which is obtained upon making use of the covariant self-conservation of the effective DE fluid,

\begin{equation}\label{eq:fDE}
       \rho_{{\rm DE},J}(a)=\rho_{{\rm DE},J}^0\,a^{-3\left(1+{\sum\limits_{j=0}^{J}w_j}\right)}\exp \left[ -3 \sum_{l=1}^{J} \sum_{k=1}^{l} \binom{l}{k}(-1)^{k}\frac{ w_{l}}{k}(a^{k}-1)\right]\,.
\end{equation}
In the $\rho$-basis, instead, the DE density reads,

\begin{equation}\label{eq:rhofam}
\rho_{\mathrm{DE},J}(a) =\rho_{{\rm DE},J}^0 \left[1+\sum_{l=1}^{J}C_l(1-a)^l\right]\, ,
\end{equation}
with the associated EoS parameter,

\begin{equation}\label{eq:B2w}
    w_{{\rm DE},J}(a) = -1+\frac{a}{3}\left[\frac{{\sum\limits_{l=1}^{J}lC_l(1-a)^{l-1}}}{1+{\sum\limits_{l=1}^{J}C_l(1-a)^{l}}}\right]\,.
\end{equation}
The number of additional parameters with respect to $\Lambda$CDM is $n_J=J+1$ and $n_J = J$ in the $w$-basis and $\rho$-basis, respectively. For the latter, this also corresponds to the maximum number of zeros that the DE density in model $J$ can exhibit in the range $a \in [0,1]$. Notice that in other WFR bases, such as the $w$-basis, negative values of the DE density in the past are not allowed if the present value is positive, as preferred by the data. It is therefore interesting that the $\rho$-basis enables the exploration of this region. In this way, and following the approach of \cite{Gonzalez-Fuentes:2025lei}, we can assess to what extent negative values of $\rho_{\mathrm{DE},J}(a)$ are required -- a feature encountered in several models in the literature and even favored in certain contexts, particularly in relation to the Hubble tension and in the presence of angular (two-dimensional, 2D) BAO data (see, e.g., \cite{Bernui:2023byc,Akarsu:2023mfb,Dwivedi:2024okk,Gomez-Valent:2024ejh,Gomez-Valent:2024tdb,Xu:2026sbw,Akarsu:2026anp}).

As a guiding principle, we take as many models within a basis as necessary to surpass the peak in the weights. As shown in Tables \ref{tab:tab_des_w} and \ref{tab:tab_des}, for DES-Dovekie it is sufficient to consider 3 models in both bases, whereas with Pantheon+ we are forced to keep one more model, so in this case we only employ the $\rho$-basis (cf. Table \ref{tab:tab_panplus}).  For all the models used for the reconstruction, we have checked that the likelihood is Gaussian in good approximation.\\  

\begin{table*}
\begin{center}
\begin{tabular}{|c|cccc|c|}
\multicolumn{6}{c}{\textbf{CMB + DESI DR2 + PantheonPlus, $\rho-$basis}}\\ \hline
Param. & $n_J=1$ & $n_J=2$ & $n_J=3$ & $n_J=4$ & Reco.\\ \hline
$H_0$& $67.48\pm0.60$ & $67.43\pm 0.61$ & $67.75\pm0.62$ & $67.52\pm0.67$ & $67.62\pm 0.64$\\
$10^2\omega_{\rm b}$ & $2.246\pm 0.013$ & $2.236\pm0.013$ & $2.227\pm 0.014$ & $2.224\pm 0.014$ & $2.228\pm 0.015$\\
$10\omega_{\rm cdm}$ & $1.166\pm 0.008$ & $1.181\pm0.010$ & $1.193\pm 0.012$ & $1.196\pm0.012$ & $1.191\pm 0.013$\\
$C_1$ & $0.13\pm 0.10$ & $0.56\pm0.23$ & $-0.24\pm 0.45$ & $0.53\pm 0.86$& $-$ \\
$C_2$ & $-$ & $-1.15\pm0.53$ & $3.45\pm 2.34$ & $-3.79\pm 7.24$ & $-$\\
$C_3$ & $-$ & $-$ & $-5.86\pm2.87$ & $14.44\pm 19.32$ & $-$ \\
$C_4$ & $-$ & $-$  & $-$ & $-16.81\pm15.76$ & $-$\\ \hline
$\Omega_{\rm m}^0$ & $0.307\pm 0.006$ & $0.310\pm 0.006$ & $0.310\pm 0.006$& $0.313\pm 0.007$ & $0.311\pm 0.006$\\
$F_\mathrm{cross}\left[\%\right]$ & $0.00$ & $97.58$ & $99.89$ & $99.97$ & $96.73$\\ 
$F_\mathrm{neg}\left[\%\right]$ & $0.00$ & $0.00$ & $13.02$ & $10.26$ & $9.63$\\
$\chi^2_{\rm min}$ & $1420.92$  & $1415.97$ & $1411.71$ & $1411.27$ & $-$\\
$p-$value & $0.183$ & $0.0346$ & $ 0.0118$ & $0.0222$ & $-$ \\
$W_J\left[\%\right]$ & $2.78$  & $17.40$  & $52.30$ & $27.52$ & $-$\\ \hline
\end{tabular}
\caption{Same as in Table \ref{tab:tab_des}, but using CMB $+$ DESI DR2 $+$ PantheonPlus in the $\rho$-basis. $H_0$ is given in km/s/Mpc.}\label{tab:tab_panplus}
\end{center}
\end{table*}

\begin{figure}[t!]
    \centering
    \includegraphics[scale=0.52]{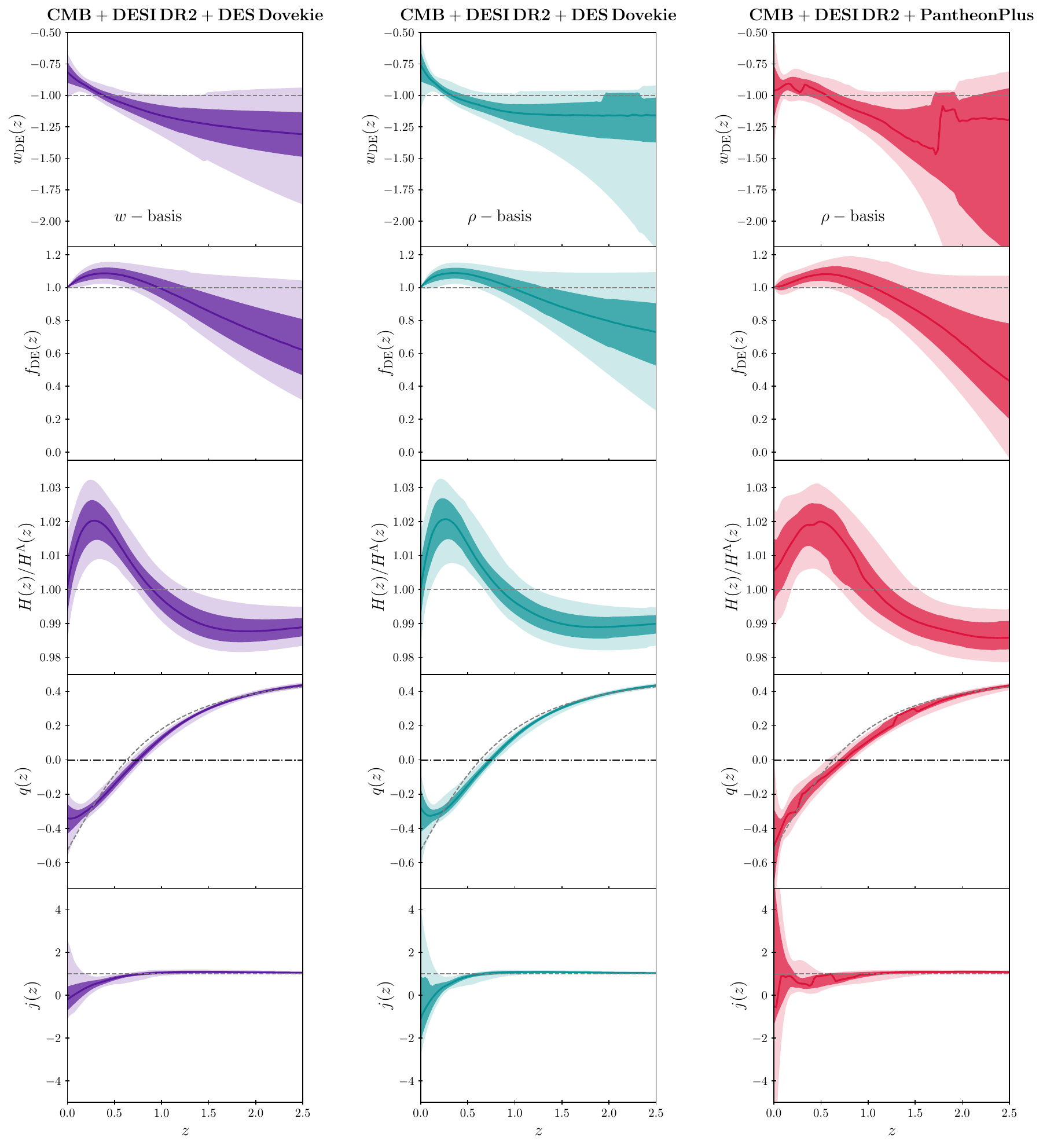}
    \caption{Reconstruction of the EoS parameter, normalized DE density, Hubble rate, deceleration parameter and jerk obtained with the datasets and WFR-bases analyzed in section \ref{sec:standardphysics}. For the datasets containing DES-Dovekie we have used $n_J=3$ in both $w-$ and $\rho-$basis, whereas for Pantheon+ we have employed the $\rho-$basis with $n_J=4$, see the main text for details. We show the Hubble rate  normalized to the \textit{Planck}PR4 best-fit $\Lambda$CDM model, which sets  $\Omega_{\rm m}^0 = 0.315$ and $H_0=67.26\,$km/s/Mpc \cite{Rosenberg:2022sdy}. In each of the plots, gray dashed lines correspond to $\Lambda$CDM and the black dash-dotted line in the $q(z)$ plot sets the border between deceleration and acceleration. }
    \label{fig:reco_standard}
\end{figure}

In our Tables \ref{tab:tab_des_w}-\ref{tab:tab_panplus}, apart from displaying the fitting results of the individual models, we also provide the probability of phantom crossing $F_\mathrm{cross}$ from $w_{\rm DE}<-1$ (at $z>z_c$) to $w_{\rm DE}>-1$ (at $z<z_c$) and the probability of having negative energy density in the range $0\leq z \leq 3$, $F_\mathrm{neg}$. The evidence for phantom crossing seems to be a stable signal across the different bases and SNIa samples: $98.46\%$ ($96.86\%$) for DES-Dovekie in the $w$ ($\rho$) $-$basis and $96.73\%$ for Pantheon$+$. These numbers are consistent with those reported in \cite{Gonzalez-Fuentes:2025lei}, and lie also in the ballpark of the values obtained with other methods, see, e.g., \cite{Keeley:2025rlg}. On the contrary, the probability of having negative energy density in $z\in [0,3]$ with Pantheon$+$ is significantly higher than the corresponding one in Tables 5 and 6 of \cite{Gonzalez-Fuentes:2025lei}. We consider this discrepancy a consequence of the inherent uncertainty in obtaining $F_{\rm neg}$. As discussed in our previous work, the last BAO point is at the effective redshift $z_{\rm eff}=2.330$ and for $z\gtrsim3$ the shape of $f_\mathrm{DE}$ cannot be trusted. Since the redshift at which our reconstructions exhibit negative energy densities are close to this region, it is hard to give a reliable quantification of $F_{\rm neg}$ with current data. Notice that this is in contradistinction with $F_{\rm cross}$, which is a stable feature at a redshift where the reconstructions can be regarded as trustworthy. 

The plots of the reconstructed functions are presented in Fig. \ref{fig:reco_standard}. For the three combinations of datasets and bases, we display the EoS parameter, the normalized DE density and Hubble rate, the deceleration parameter ($q=-\ddot{a}/(aH^2)$, with the dot denoting a derivative with respect to cosmic time), and the jerk ($j=\dddot{a}/(aH^3)$). Using the SNIa sample from DES-Dovekie, we observe an overall reduction in the uncertainties compared to those obtained with Pantheon+. The significance of the peak in the effective DE density is slightly enhanced. This is consistent with the decrease in the $p-$values obtained with the FB method for the individual models of the $\rho$-basis when using DES-Dovekie, cf. again Tables \ref{tab:tab_des} and \ref{tab:tab_panplus}. In addition, it is important to highlight that the comparison of the reconstructed functions in the first and second columns of Fig. \ref{fig:reco_standard}, which contain the results obtained using CMB $+$ DESI DR2 $+$ DES Dovekie with the $w$- and $\rho$-basis, respectively, make it clear that the constraints on the various physical quantities are fully compatible, meaning that the choice of the basis does not have a major impact on our results. This conclusion is in perfect agreement with our previous work \cite{Gonzalez-Fuentes:2025lei}, in which we employed information criteria (AIC and DIC) for the computation of the weights. We see that now, under this more rigorous approach based on the FB method, the results remain stable.


\subsection{DE reconstruction with new physics before recombination}\label{sec:newphysics}

In this section, we avoid using the CMB information derived under the assumption of standard pre-recombination physics. To remain as model-agnostic as possible and obtain results largely independent of the details of the underlying physics governing the universe prior to decoupling, our fitting analysis only incorporates the location of the first CMB acoustic peak through the corresponding prior on $\theta_* = r_*/D_M(z_*)$, as described in section \ref{sec:data}. We exploit the tight relation between the sound horizon at decoupling ($r_*$) and at the baryon-drag epoch ($r_d$), namely $r_*= 0.983\,r_d$, which holds to very good approximation across essentially all models, regardless of the specific pre-decoupling physics they introduce \cite{Sharma:2025iux}. We therefore sample over $r_d$ and compute $r_*$ using this relation. Instead of allowing the Boltzmann code to determine $z_*$ as the maximum of the visibility function within a standard recombination scenario, we fix it to the value obtained in the CPL best-fit model with CMB$+$DESI DR2$+$PantheonPlus from section \ref{sec:FBresults}, $z_* = 1088.71$. Our results can nevertheless be straightforwardly extrapolated to other values of $z_*$, since the comoving distance $D_M(z_*)$ depends only weakly on this redshift: it enters solely as the upper limit of the integral, and treating it as a free parameter would leave it largely unconstrained, as we have explicitly verified.

\begin{figure}
    \centering
    \includegraphics[width=\linewidth]{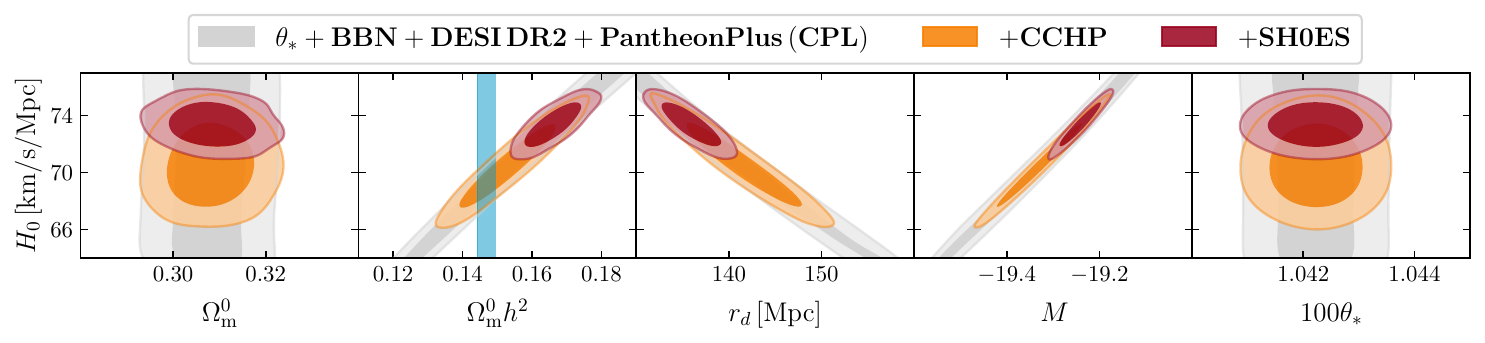}
    \caption{Contour plots obtained for the CPL parametrization making use of the uncalibrated combination of data and the corresponding ones obtained upon applying the calibration from SH0ES or CCHP. The corresponding fitting results are displayed in Tables \ref{tab:tab_sh0es_w} and \ref{tab:tab_cchp_w}, respectively. In the plane $(\omega_m,H_0)$, we also display in a blue vertical band the $1\sigma$ EDE constraint $\omega_m=0.1468\pm0.0026$  from the profile likelihood analysis of Ref. \cite{Gomez-Valent:2022hkb}, obtained with the full {\it Planck} 2018 likelihood in combination with SNIa and BAO, and without including the SH0ES prior. This is to illustrate the existing tension between the required values of $\omega_m$ that would be needed to explain the SH0ES measurement in scenarios with new physics before the decoupling and the one obtained from a full CMB analysis. Similar constraints on $\omega_m$ are also obtained in models that speed up the recombination process, see e.g. Table II of \cite{Mirpoorian:2025rfp}.}
    \label{fig:corrrelations_calib}
\end{figure}

The rationale behind this approach is to remain model-agnostic with respect to the new physics introduced prior to recombination to reduce the sound horizon. The model-independent constraints obtained in this section arise from the combination of SNIa, BAO, and $\theta_*$ (which can be interpreted as an additional BAO point at $z_*$), together with the calibrators. The uncalibrated dataset alone is sufficient to constrain the normalized Hubble rate, $E(z)=H(z)/H_0$, particularly at late times. It places tight constraints on $\Omega_{\rm m}^0$ and the DE parameters ($w_0, w_a, \ldots$), while leaving $r_d$, $H_0$, and $M$ unconstrained \cite{Lin:2021sfs,Poulin:2024ken,Sharma:2025iux, Wang:2025mqz}. We illustrate this in Fig. \ref{fig:corrrelations_calib} (see the gray bands therein), using the CPL parametrization in the late universe.

Once a calibrator is included -- e.g., distances to host galaxies from SH0ES, which calibrate the absolute magnitude of SNIa, or independent measurements of $H_0$, or full CMB information within a specific recombination scenario -- the degeneracies in the $r_d$–$h$–$M$ planes of parameter space are lifted. The BBN prior on $\omega_{\rm b}$ breaks the degeneracy between baryonic and cold dark matter, but has no further impact on our analysis, since the rest of cosmological observables employed are not sensitive to the splitting of $\omega_m$ in $\omega_{\rm b}$ and $\omega_{\rm cdm}$. If the calibrator anchors $H_0$ to high values, this may lead to a very large matter density, $\omega_m$ \cite{Jedamzik:2020zmd,Vagnozzi:2023nrq,Poulin:2024ken,Pedrotti:2024kpn}\footnote{Models that introduce new physics prior to decoupling but after BBN in order to alleviate the Hubble tension typically require values of $\omega_{\rm b}$ larger than those preferred by BBN constraints once the full CMB likelihood is considered. This leads to a tension in $\omega_{\rm b}$ \cite{Giovanetti:2026aku}. In our model-independent approach, we cannot make use of the full CMB information and are therefore not sensitive to this tension. However, there remains a strong tension in $\omega_{\rm m}$, which is largely independent of the former.}. This, however, does not affect the late-time normalized expansion history, and therefore the constraints on $\Omega_{\rm m}^0$, $w_0$, $w_a$, etc., remain stable \cite{Sharma:2025iux}. In Fig. \ref{fig:corrrelations_calib} (see also the corresponding CPL results displayed in Table \ref{tab:tab_sh0es_w}), we show that in order to explain comfortably values of the Hubble parameter in the ballpark of that measured by SH0ES, the underlying model should be able to accommodate values of $\omega_m\gtrsim 0.16$, which are clearly in tension with those inferred from fitting analyses that do not include the SH0ES prior and incorporate the full CMB information from {\it Planck} (cf. the vertical blue band and the red contours in the $\omega_m-H_0$ plane of Fig. \ref{fig:corrrelations_calib}). They could lead to larger amplitudes of the matter power spectrum\footnote{However, we should keep in mind that the growth of density fluctuations is controlled by $\Omega_{\rm m}(a)$ (not $\omega_m$), so it is not trivial to see how redshift-space distortions and weak lensing measurements will be affected, also because here we are not fitting the parameters of the primordial power spectrum. An increase of $\omega_m$ does not necessarily lead to an increase of $P(k)$; see, e.g., the discussion in the context of EDE of Ref. \cite{Gomez-Valent:2022hkb}.}. It is important to note that this tension in $\omega_m$ exists regardless of the concrete scenario that one uses to decrease $r_d$. It is found both in models that inject energy before the decoupling and in models that accelerate the recombination process, see, e.g., \cite{Gomez-Valent:2022hkb,Mirpoorian:2025rfp}. This may call into question the ability of these models to explain the SH0ES measurement, even when aided by the additional freedom provided by late-time dynamical dark energy scenarios.

\begin{table*}[t!]
\begin{center}
\begin{tabular}{|c|cc|c|}
\multicolumn{4}{c}{$\mathbf{\theta_*}$ \textbf{+ BBN + DESI DR2 + PantheonPlus + SH0ES, $w-$basis}}\\ \hline
Parameter & $w$CDM & CPL & Reconstructed  \\ \hline
$H_0$[km/s/Mpc] & $73.5\pm 1.0$ & $73.3\pm 1.0$ & $73.4\pm 1.0$\\
$10^2\omega_{\rm b}$ & $2.196\pm 0.063$ & $2.196\pm 0.063$ & $2.196\pm0.063$\\
$10\omega_{\rm cdm}$ & $1.418\pm 0.052$ & $1.430\pm 0.053$ & $1.422\pm 0.052$\\
$r_d$ [Mpc]& $136.1\pm 2.1$ & $136.0\pm 2.1$ & $136.1\pm2.1$ \\
$M$ & $-19.245\pm0.030$ & $-19.244\pm 0.030$ & $-19.245\pm 0.029$\\
$w_0$ & $-0.944\pm 0.027$ & $-0.889\pm 0.060$ & $-$\\
$w_a$ & $-$ & $-0.28\pm 0.27$ & $-$\\ \hline
$\Omega_{\rm m}^0$ & $0.304\pm 0.005$ & $0.308\pm0.006$ & $0.306^{+0.005}_{-0.006}$\\
$F_\mathrm{cross}\left[\%\right]$ & $0.00$ & $78.10$ & $29.25$ \\
$\chi^2_{\rm min}$ & $1461.78$  & $1460.52$ & $-$\\
$p-$value & $0.03879$ & $0.06313$ & $-$\\
$W_J\left[\%\right]$ & $62.54$ & $37.46$ & $-$\\ \hline
\end{tabular}
\caption{Same as in Table \ref{tab:tab_des_w}, but using the priors in $\theta_*+$BBN, together with the BAO data from DESI DR2 and the Cepheid-calibrated SNIa from PantheonPlus+SH0ES. We employ the $w-$basis. For this dataset, $\Lambda$CDM yields $\chi^2_{\rm min} = 1466.05$.}\label{tab:tab_sh0es_w}
\end{center}
\end{table*}

We apply now the WFR method to reconstruct the dark energy and the main cosmographical functions in order to learn about them in a more controlled setup that minimizes the impact of subjective choices in the analysis. The fitting results for the individual models contributing to the weighted reconstruction are displayed in Tables \ref{tab:tab_sh0es_w} and \ref{tab:tab_cchp_w} for the cases in which we calibrate the ladders with SH0ES and CCHP, respectively. We use the $w$-basis, and see that the data under consideration do not require a higher degree of complexity than the one encountered in the CPL. The maximum weight is actually obtained for the $w$CDM. We therefore use these two parametrizations in our reconstructions, since going beyond the CPL would probably incur into overfitting.

Our WFR reconstruction allows us to determine the conditions that must be fulfilled by any model that aims to solve the $H_0$ tension, while keeping a good description of the low-$z$ data, and quantify the evidence of dynamical DE in this scenario. We stress that the reconstructions from this section cannot be compared on equal-footing with those from section \ref{sec:standardphysics}. Here we are not using any correlation nor CMB constraint involving the total pressureless matter density, since they can change in different pre-recombination scenarios and we want to keep the analysis general to allow for more model-independent conclusions. 

We first comment on the fact that in our reconstructions with non-standard physics before recombination we find very similar values of $\chi^2_{\rm DESI}$ and $\chi^2_{\rm PanPlus}$ to those found in section \ref{sec:standardphysics} with the model with minimum $\chi^2$ \footnote{In order to perform this comparison on equal footing, we have repeated the fitting analysis of section \ref{sec:standardphysics} without marginalizing over the absolute magnitude, and in the case of the analysis with SH0ES we have computed $\chi^2_{\rm PanPlus}$ without considering the contribution of the distance moduli to the host galaxies.}, meaning that the goodness-of-fit to the low-redshift data is fully comparable. In other words, we find no significant improvement when we allow $r_d$ to vary freely in the fitting analysis, regardless of the calibrator employed.

Using both the SH0ES and CCHP calibrations, we find a mild preference for dynamical DE in the form of quintessence, at the $\sim 2\sigma$ confidence level for $z < 0.3$. At higher redshifts, this preference further weakens, and $w_{\rm DE}(z)$ becomes compatible with $-1$ at the $\sim 1\sigma$ level. While a crossing of the phantom divide is not excluded, there is no compelling evidence for it. Notably, had we only employed the $w$CDM model to study the DE dynamics, we would have concluded that the EoS parameter lies in the quintessence regime for all $z$, at $2.1\sigma$ confidence. This highlights the importance of including the CPL parametrization to obtain a more realistic reconstruction and, consequently, more robust conclusions.

\begin{table*}[t!]
\begin{center}
\begin{tabular}{|c|cc|c|}
\multicolumn{4}{c}{$\mathbf{\theta_*}$ \textbf{+ BBN + DESI DR2 + PantheonPlus + CCHP, $w-$basis}}\\ \hline
Parameter & $w$CDM & CPL & Reconstructed  \\ \hline
$H_0$[km/s/Mpc] & $70.3\pm 1.9$ & $70.3\pm 1.9$ & $70.3\pm 1.9$\\
$10^2\omega_{\rm b}$ & $2.196\pm 0.063$ & $2.196\pm 0.063$ & $2.196\pm 0.063$ \\
$10\omega_{\rm cdm}$ & $1.275\pm 0.088$ & $1.295\pm 0.091$ & $1.282\pm 0.089$ \\
$r_d$ [Mpc]& $142.3\pm 4.1$ & $141.9\pm 4.1$ & $142.2\pm 4.1$\\
$M$ & $-19.343\pm 0.060$ & $-19.336\pm 0.061$ & $-19.340\pm 0.060$\\
$w_0$ & $-0.943\pm 0.027$ & $-0.889\pm 0.060$ & $-$\\
$w_a$ & $-$ & $-0.28\pm 0.27$ & $-$\\ \hline
$\Omega_{\rm m}^0$ & $0.304\pm 0.005$ & $0.308\pm 0.006$ & $0.305_{-0.006}^{+0.005}$\\
$F_\mathrm{cross}\left[\%\right]$ & $0.00$ & $76.99$ & $27.95$ \\
$\chi^2_{\rm min}$ & $1412.52$ & $1411.33$ & $-$\\
$p-$value & $0.03443$ & $0.05889$ & $-$\\
$W_J\left[\%\right]$ & $63.70$ & $36.30$ & $-$\\ \hline
\end{tabular}
\caption{Same as in Table \ref{tab:tab_sh0es_w}, but using the local measurement from TRGB calibration $H_0^{\rm CCHP}$ \cite{Freedman:2024eph}, using again the $w-$basis. For this dataset, $\Lambda$CDM yields $\chi^2_{\rm min} = 1416.99 $.}\label{tab:tab_cchp_w}
\end{center}
\end{table*}

\begin{figure}[t!]
    \centering
    \includegraphics[width=0.8\linewidth]{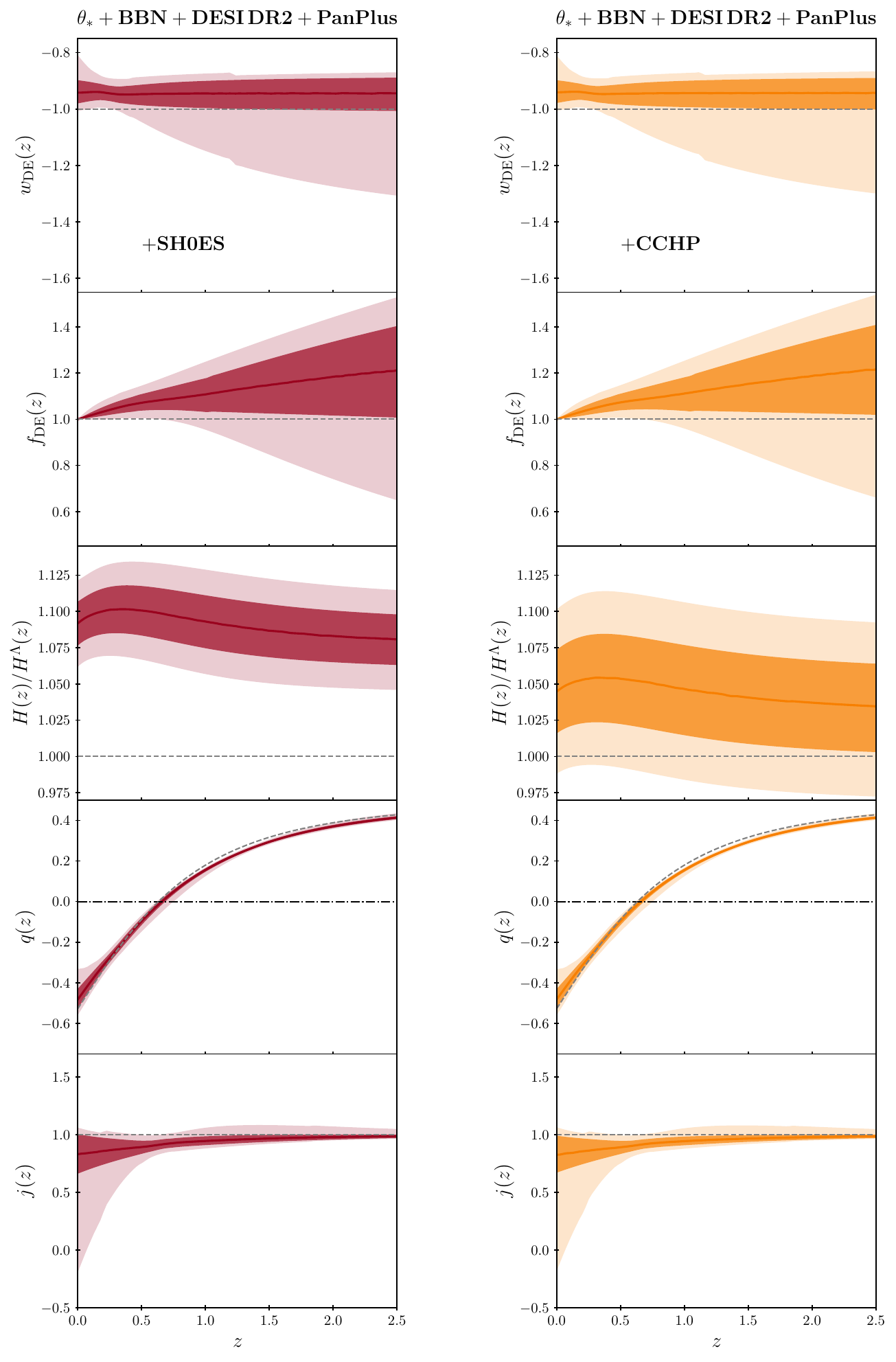}
    \caption{WFR-reconstructions of the DE background properties and the first three cosmographical functions using the $w$-basis and allowing deviations from standard physics before recombination. We employ the dataset $\theta_*$+BBN+DESI DR2+PantheonPlus calibrated with the measurements from SH0ES (in red) or CCHP (in yellow). The Hubble function $H(z)$, has been normalized with the {\it Planck}PR4 best-fit $\Lambda$CDM model, as in Fig. \ref{fig:reco_standard}. See the comments in the main text.}\label{fig:reco_newPhys}
\end{figure}

But what is causing such a reduction in the evidence for dynamical DE? From the results displayed in our Tables \ref{tab:tab_sh0es_w} and \ref{tab:tab_cchp_w} we see that our constraints on the DE parameters obtained for the individual models lead to weaker evidences for dynamical dark energy than those reported in section \ref{sec:standardphysics}, which is is in full agreement with our reconstructions, of course. Thus, one might be tempted to conclude that the evidence for dynamical DE is reduced when the calibration from SH0ES is employed. However, it is important to note, first, that this decrease is already present when using uncalibrated data. Indeed, our central values of the DE parameters are close to those obtained by DESI within the CPL framework using the DESI+PantheonPlus data alone (with no CMB information nor ladders calibration), cf. Table V of \cite{DESI:2025zgx}.   The inclusion of $\theta_*$ does not induce significant parameter shifts -- although it does provide additional constraining power. On the other hand, the breaking of the $r_d$–$H_0$ degeneracy alone does not significantly affect the level of evidence of dynamical DE. One can understand that by comparing the constraints on the DE parameters obtained with the SH0ES and CCHP calibrations. The aforesaid constraints are extremely similar, even though the two calibrations differ both in the preferred absolute value of $H_0$ and its uncertainty. This is a reflection of the insensitivity of the DE parameters to the specific calibration used in the analysis. Therefore, the decrease in significance of the preference for dynamical dark energy arises primarily from discarding the additional information provided by the CMB under the assumption of standard recombination. This result resonates well with that of Ref. \cite{Wang:2025mqz}.

Even in this conservative approach\footnote{Which is also optimistic in the sense that we are taking for granted that the model that reduces $r_d$ does not spoil the rest of CMB data.} -- where no specific model of early-time new physics is assumed -- the fact that $\Omega_{\rm m}^0$ is well determined by late-time observables implies an increase in $\Omega_{\rm m}^0 h^2$ when early-Universe physics is invoked to reduce the sound horizon with the aim of explaining the values of $H_0$ preferred by SH0ES or CCHP\footnote{This increase in the total pressureless matter density is also accompanied by an increase in the effective DE density in order to keep the value of $\Omega_{\rm m}^0$ stable (cf. again Fig. \ref{fig:reco_newPhys}).}. This provides further motivation for a model-agnostic analysis, since using the full CMB likelihood can induce non-trivial parameter shifts that may appear to improve the fit despite causing internal inconsistencies between the various datasets involved. For instance, EDE models are known to alleviate the $H_0$ tension. However, when constrained by the full CMB likelihood, they typically reach only $H_0 \sim 70\,\mathrm{km s^{-1}Mpc^{-1}}$ (even in profile likelihood analyses), while requiring relatively large values of $\omega_m \sim 0.147$ to compensate for the impact of EDE on the early integrated Sachs–Wolfe effect \cite{Vagnozzi:2021gjh}. A complete resolution of the Hubble tension would further exacerbate these high $\omega_m$ values, whereas joint fits including the full CMB likelihoods tend to dilute the tension in $\omega_m$ at the expense of worsening the one in  $H_0$.

The deceleration parameter also exhibits some interesting features. We find that in these scenarios, in the redshift range $z\in(0.5,2)$, there is a preference for a less decelerated universe than in the best-fit $\Lambda$CDM model from {\it Planck} PR4, which could also have a non-negligible impact on the large-scale structure formation processes in the universe.  

Of course, as we have already mentioned above, these reconstructions rely very heavily on the ability of the model to explain larger values of $\omega_m$ and, at this point, it is important to remark that there is no model in the literature capable of doing so at the level required by SH0ES. However, intermediate values of $H_0$, closer to 70 km/s/Mpc, as those measured by CCHP, could be still accommodated by well-known models as EDE or by the presence of primordial magnetic fields before the decoupling of the CMB photons.



\section{Conclusions}\label{sec:conclusions}

Current cosmological data provide our only window into the properties of dark energy\footnote{Aside from local gravity probes that can constrain interactions between dark energy and visible matter with Earth, near-Earth or Solar System tests, see, e.g., \cite{Brax:2018zfb,Elder:2019yyp,Vagnozzi:2021quy,Tsai:2021irw,March:2021mqu,Brax:2022olf,Tsai:2023zza,Feleppa:2025clx,Feleppa:2025vop,Yuan:2025twx}.}. In our effort to characterize its main features as a step toward a more fundamental understanding \cite{SolaPeracaula:2022hpd}, a natural question is whether it evolves with the cosmic expansion or it is, instead, a rigid entity. Although hints of dark energy dynamics were already found in different contexts in the past \cite{Alam:2003fg,Alam:2004jy,Salvatelli:2014zta,Sahni:2014ooa,Sola:2015wwa,Sola:2016jky,SolaPeracaula:2016qlq,Zhao:2017cud,SolaPeracaula:2017esw,Sola:2017znb,SolaPeracaula:2018wwm}, the BAO data recently reported by DESI and the state-of-the-art data from supernovae of Type Ia and CMB show a preference not only for an evolving dark energy density, but also for a crossing of the phantom divide happening at $z_c\sim 0.5$. There is, however, considerable debate within the community regarding the actual significance of these hints, with Bayesian analyses (based on the computation of Bayes ratios) finding no evidence for dark energy evolution and with frequentist pipelines leading to a signal at $\sim 3\sigma$ CL. Nevertheless, the conclusions extracted from the Bayesian analyses are at odds with the marginalized constraints in the $w_0-w_a$ plane of the CPL, for instance. In addition, the computation of Bayes ratios depends strongly on the size of the prior volume, even in cases where priors are highly uninformative and play no role in determining the shape of the posterior distribution.

In this work, with the aim of mitigating these problems and clarifying the situation, we have applied the Frequentist-Bayesian (FB) method developed in Refs. \cite{Keeley:2021dmx,Amendola:2024prl}, in which the Bayes factor is treated as a frequentist random variable. We have re-analyzed the CPL parametrization in the light of current CMB+BAO+SNIa data, considering both the SNIa from Pantheon+ and the re-calibrated sample from DES, i.e., DES-Dovekie, which corrects some issues that were afflicting the previous DES-Y5 sample. Our determination of the significance of the preference for dynamical dark energy is in full agreement with the frequentist method. We find a $\sim 2.5\sigma$ hint with Pantheon+, which is enhanced to the $\sim 3\sigma$ CL with DES-Dovekie.

Then, we have applied the Weighted Function Regression (WFR) method to reconstruct the most important background dark energy properties and cosmographical functions, mitigating the impact of parametrization-dependent features on the results and under the assumption of standard pre-recombination phyics. We improve the computation of the weights with respect to the one applied in our previous work \cite{Gonzalez-Fuentes:2025lei}. Now, we base it also on the FB method, instead of relying on information criteria. This can be efficiently done by exploiting the Fisher matrix formalism and the availability of an analytical expression for the distribution of the Bayes factors. Our analysis shows that the probability of crossing of the phantom divide remains stable with respect to that reported in \cite{Gonzalez-Fuentes:2025lei}, although it is somewhat smaller than that obtained with DES-Y5. We find the probability of crossing to be $\sim 96.7 \-- 98.5\%$.

Since these reconstructions lead still to values of $H_0$ closer to the {\it Planck}/$\Lambda$CDM value and, hence, in stark tension with the SH0ES measurement, we deem interesting to reconstruct the low-$z$ history of the universe, but this time remaining fully agnostic regarding the new possible physics entering before the decoupling of CMB photons and calibrating the cosmic ladders either with the distances to the host galaxies from SH0ES or the TRGB measurement from CCHP. One of our main results is that any model aiming to explain the SH0ES measurement should be able to accommodate a very large value of the reduced matter density parameter, typically $\omega_m>0.16$, which is clearly in conflict with the values obtained from full CMB analysis. This result is in agreement with that reported in previous works \cite{Jedamzik:2020zmd,Poulin:2024ken,Pedrotti:2024kpn}. We remark that virtually no model available in the literature can do that job efficiently. The CCHP calibration gives still margin for early-time resolutions to the Hubble tension, though. 

The late-time reconstruction of the dark energy density in this case shows no significant evidence for crossing of the phantom divide, although such a crossing is not excluded. The largest deviation from $\Lambda$CDM is found at $z \lesssim 0.3$, at most at the $2\sigma$ CL. However, these reconstructions rely on the model’s ability to accommodate larger values of $\omega_m$, which may not be achievable within known pre-recombination modifications. This does not imply that models reducing the sound horizon cannot resolve the $H_0$ tension; rather, it suggests that reducing the sound horizon alone may be insufficient. We have explicitly shown that this remains the case even in a model-agnostic reconstruction of dark energy behavior at late times.


\acknowledgments AGF is supported by grant FPU24/01241 (MICIU). AGV is funded by “la Caixa” Foundation (ID 100010434) and the European Union's Horizon 2020 research and innovation programme under the Marie Sklodowska-Curie grant agreement No 847648, with fellowship code LCF/BQ/PI23/11970027. He is also supported by projects PID2022-136224NB-C21 (MICIU), 2021-SGR-00249 (Generalitat de Catalunya) and CEX2024-001451-M (ICCUB). Both authors acknowledge the participation in the COST Action CA21136 “Addressing observational tensions in cosmology with systematics and fundamental physics” (CosmoVerse).


\bibliographystyle{JHEP}
\bibliography{biblio.bib}

\end{document}